# PHOTOEXCITED CARRIERS RECOMBINATION AND TRAPPING IN SPHERICAL VS FACETED TiO$_2$ NANOPARTICLES

Gianluca Fazio, Lara Ferrighi, Cristiana Di Valentin[*]

Dipartimento di Scienza dei Materiali, Università di Milano-Bicocca, via R. Cozzi 55, 20125 Milano Italy

[*]Corresponding author: cristiana.divalentin@mater.unimib.it

**Abstract**

Nanoparticles of very small size (below 10 nm) of TiO$_2$ material are nowadays the functional building blocks of many developing technological applications. Nano is clearly different from bulk or extended systems as regards surface area, molecular binding properties, charge separation efficiency, electron/hole transport, photochemical conversion properties, etc. In this work, we investigate the life path of energy (excitons) and charge (electrons and holes) carriers in anatase TiO$_2$ nanoparticles of different size (2-3 nm) and shape (faceted vs spherical), by means of a wide set of hybrid density functional theory calculations. The attention is focused on the exciton/charge carriers formation, separation, recombination, self-trapping processes, which are analyzed in terms of structural deformations, energy gain or cost, charge localization/delocalization and electronic transitions involved. The computational models are corroborated by an extensive comparison with available experimental data based on photoluminescence measurements, electron paramagnetic resonance and transient absorption spectroscopies. Peculiar differences are observed for spherical nanoparticles with respect to faceted ones because of the higher disorder and larger diversity of coordination sites present on the surface. For example, charge delocalization on several lattice sites is more competitive with self-trapping processes in faceted than in spherical nanoparticles. This relates to the fact that selective compression or elongation of Ti-O bonds play a key role in determining the effectiveness of trapping sites, with spherical nanoparticles being more flexible. Moreover, hydroxyl groups on surface five-fold coordinated Ti sites are also found to be good hole trapping sites.

Keywords: nanoparticles, photochemistry, photocatalysis, photovoltaics

## 1. Introduction

Many technological applications make use of titanium dioxide nanoparticles and of their ability to convert light photons energy into chemical processes for efficient catalysis and electrochemistry.[1,2,3,4] In this respect, photocatalysis,[2,3] photoelectrochemistry[5] and photovoltaics[1] are modern and fast-developing research fields where titanium dioxide is still considered a "superior" material.

The ability of $TiO_2$ materials to interact with light and produce photoinduced charge carriers is still a fascinating and challenging area of research, especially when effects due to the nanosized dimensions come into play.[6] Understanding the processes at the basis of the irradiated light energy conversion into chemical species with intrinsic redox potential is extremely important. In the limited volume of a nanoparticle, these processes may be characterized by shorter or different timescales leading to peculiar effects of high interest for the research community involved in the study and application of nanomaterials.[7]

When the interaction between light and a semiconductor material takes place, an exciton is initially formed.[8,9] The dimension of the exciton largely depends on the dielectric constant of the material and the effective masses of the excited electron and hole.[10] When strongly coupling with lattice vibrations, excitons may become self-trapped,[11] which largely reduces their mobility. The physical confinement of an exciton in a nanoparticle of few nanometers size may significantly affect its size and localization.[7] Moreover, it is still an open and critical issue whether the close presence of borders, e.g. facets, edges and corners, may accelerate the process of excitons separation into charge carriers of opposite charge (electrons and holes) or, on the contrary and for different reasons, may accelerate the electron/hole recombination through exciton self-trapping processes.[12]

The accurate theoretical description of all these aspects requires the recourse to sophisticated ab initio methods. Density functional theory (DFT) could be a viable approach, providing a good balance between accuracy and computational cost. Realistic nanoparticles of few nanometers size contain a rather high number of atoms. For example, the smallest nanoparticles with a well-defined crystalline phase contain at least 500 atoms, with an average diameter of about 2 nm.[13,14] However, the electron self-interaction problem inherent in DFT is a severe issue for the description of self-trapped excitons or self-trapped electrons and holes.[15,16,17,18] For this reason, some self-interaction correction is required as it is present in the more costly hybrid density functional methods.

Additionally, besides the size of the nanoparticle, also the shape can play a relevant role.[19] Faceted nanocrystals are the most thermodynamically stable,[20] as predicted with DFT calculations in terms of standard Wulff construction.[21] Under specific experimental conditions, such as excessive dilution, spherical nanoparticles are formed, which are particularly rich of undercoordinated sites.[22] These are expected to play a key role in the self-trapping processes of charge carriers. However, it is still an open question whether there is a driving force for electrons and holes to be trapped at the surface of a nanoparticle, rather than being fully delocalized in the whole nanoparticle volume. Additionally, one may expect that the exciton dynamics can be rather different in spherical with respect to faceted nanoparticles where most of the surface is made up by flat facets.

In the present computational study we follow the life path of light-induced energy carriers (excitons) or charge carriers (electrons and holes) in anatase $TiO_2$ nanoparticles, trying to provide a coherent and consistent picture of the multi-step process. We investigate, in chronological order, the exciton formation, the photogenerated charge carriers recombination and their trapping in spherical vs faceted nanoparticles, by means of a wide set of hybrid density functional calculations on nanoparticles models of increasing size and diverse shape. Various relevant quantities are computed, such as exciton self-trapping energies, recombination emission energies, energy gain associated to charge carrier separation, electron and hole self-trapping energies, electronic transitions of self-trapped species, etc. All these quantities are directly compared with the experimental findings on real $TiO_2$ nanoparticles of analogous size, as reported in the literature. These experimental data are mainly based on photoluminescence (PL) measurements, electron paramagnetic resonance (EPR) and transient absorption (TA) spectroscopy.

The manuscript is organized in two main sections, where first excitons and then separated charged carriers (electrons and holes) are discussed, respectively. Each section is introduced by an experimental background where the state-of-the-art experimental literature is briefly recalled, as the reference context for our computational results. Faceted and spherical nanoparticles are presented separately to better emphasize analogies and differences between these two types of nanosized systems.

This study highlights clear trends with size and shape, which can be rationalized in terms of structural details of the nanoparticles. In particular, distortions in the atomic structure due to the space confinement are found to be at the basis of the different ability of diverse nanoparticles to keep excitons, electrons or holes in the core or, differently, trap them at the surface. The increased presence of hydroxyls or water molecules that saturate the coordination of surface Ti ions emerges as a potential tool for improving the trapping affinity of surface species.

The aim of this work is to get a deeper insight into the processes involving energy (excitons) and charge (electrons and holes) carriers in $TiO_2$ nanoparticles of various size and shape, given that the existing literature is still mostly focused on $TiO_2$ bulk and extended surface model systems, whilst the technological applications are progressively getting more and more based on the peculiar features of $TiO_2$ nanostructured materials.

## 2. Computational Details

All the calculations were performed with the CRYSTAL14[23] package where the Kohn–Sham orbitals are expanded in Gaussian type orbitals (the all-electron basis sets are O 8-411(d1), Ti 86-411 (d41) and H 511(p1)). The B3LYP[24,25] and HSE06[26] hybrid functionals have been used throughout this work.

The values of the optimized lattice parameters are 3.789 Å and 3.766 Å for a and 9.777 Å and 9.663 Å for c, respectively for and B3LYP and HSE06, which are comparable to the experimental values.[27]

The $TiO_2$ anatase bulk was modelled by a $6\sqrt{2} \times 6\sqrt{2} \times 1$ bulk supercell with 864 atoms. To describe the (101) surface, we used a slab of ten triatomic layers with 60-atoms and a unit cell periodicity along the [10$\bar{1}$] and [010] directions; no periodic boundary conditions were imposed in the direction perpendicular to the surface. The k-space sampling for the bulk (surface) geometry optimization included a 1 1 6 (8 8 1) Monkhorst-Pack net.

Nanocrystals (**NC**) have been cut from a bulk anatase supercell following the procedure already described in our previous work.[28] Two stoichiometric nanocrystals $(TiO_2)_{159} \cdot 4\ H_2O$ (**NC$_S$**) and $(TiO_2)_{260} \cdot 6\ H_2O$ (**NC$_L$**) were carved according to the minimum energy decahedral shape,[29] where the two lowest energy anatase surface, (101) and (001), are exposed. For the chemical stability of the nanospheres (**NS**) we set for the titanium atoms a minimum of four-fold coordination and for the oxygen atoms a minimum of two-fold coordination. Thus, after carving a sphere of radius R from bulk anatase crystal, we removed all the two-fold Ti atoms on the surface, while three- and, when necessary, four-fold Ti atoms were coordinated by no more than one hydroxyl group. Oxygen atoms were saturated with a hydrogen when monocoordinated. We cut a stoichiometric nanosphere with a radius R of 1.22 nm $(TiO_2)_{223} \cdot 18\ H_2O$ (**NS**). In one case we performed calculations also on a larger sphere $(TiO_2)_{399} \cdot 32\ H_2O$ (**NS$_L$**), obtained with a radius R of 1.5 nm.

Nanoparticles have been treated as large isolated molecules in the vacuum without any periodic boundary conditions. Therefore, when an excess charge is introduced in the system, no background of charge is needed. Spin polarization is taken into account in the case of open-shell systems. Geometry optimization was performed without any symmetry constraint.

Cut-off limits in the evaluation of Coulomb and exchange series/sums appearing in the SCF equation were set to $10^{-7}$ for Coulomb overlap tolerance, $10^{-7}$ for Coulomb penetration tolerance, $10^{-7}$ for exchange overlap tolerance, $10^{-7}$ for exchange pseudo-overlap in the direct space, and $10^{-14}$ for exchange pseudo-overlap in the reciprocal space. The condition for the SCF convergence was set to $10^{-6}$ a.u. on the total energy difference between two subsequent cycles.

The gradients with respect to atomic coordinates are evaluated analytically. The equilibrium structure is determined by using a quasi-Newton algorithm with a BFGS Hessian updating scheme.[30] Convergence in the geometry optimization process is tested on the root-mean-square (RMS) and the absolute value of the largest component of both the gradients and nuclear displacements. The default thresholds for geometry optimization within the CRYSTAL code have been used for all atoms: maximum and RMS forces have been set to 0.000450 and 0.000300 a.u., and maximum and RMS atomic displacements to 0.001800 and 0.001200 a.u., respectively.

Vertical and adiabatic ionization potentials (IP) are computed by removing one electron from the model of the nanoparticle in its original geometry and by performing full atomic relaxation, respectively:

$$\mathbf{NC/NS} \rightarrow \mathbf{NC^+/NS^+} + 1\ e^- \quad (\Delta E = IP).$$

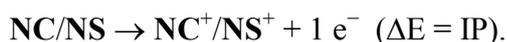

Vertical and adiabatic electron affinities (EA) are computed by adding one electron to the model of the nanoparticle in its original geometry and by performing full atomic relaxation, respectively:

$$\mathbf{NC/NS} + 1\ e^- \rightarrow \mathbf{NC^-/NS^-} \quad (\Delta E = -\ EA).$$

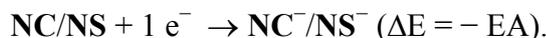

Trapping energies ($\Delta E_{trap}$) are computed as the energy difference between the isolated charges or electron-hole pairs in the trapping geometry and the delocalized solution in the neutral ground state geometry. Distortion energies ($\Delta E_{dist}$) are obtained as the difference between the energy of the neutral system in the trapping geometry and the neutral energy minimum of that nanoparticle.

The Gaussian 09 program package[31] was used to calculate EPR parameters (g, hyperfine and nuclear quadrupole tensors) for the $Ti_{29}O_{58} \cdot 4\ H_2O$ cluster with an excess hole. For these calculations an EPR-II basis set was used for O and H atoms, whereas Ti atoms were treated with a 6-311+G* basis set if first neighbor to an oxygen atom carrying the excess hole and with a 6-31G* basis set otherwise.

Simulated total densities of states (DOS) of the nanoparticles have been obtained through the convolution of Gaussian peaks ($\sigma = 0.001$ eV) centered at the Kohn-Sham energy eigenvalue of each orbital. Projected densities of states (PDOS) have been obtained by using the coefficients in the linear combination of atomic orbitals (LCAO) of each molecular orbital: summing the squares of the coefficients of all the atomic orbitals centered on a certain atom type results, after normalization, in the relative contribution of each atom type to a specific eigenstate. Then, the various projections are obtained from the convolution of Gaussian peaks with heights that are proportional to the relative contribution. For spin polarized systems, (projected) densities of states are calculated using the Kohn-Sham eigenvalues of alpha or beta population separately. The zero energy for all the DOS is set to the vacuum level, corresponding to an electron at an infinite distance from the surface.

Electronic transitions from trapping levels to the conduction band or from valence band to trapping levels are evaluated by means of the transition level approach.[32] According to Zunger et al.,[33] the optical transition between the electron trap state and the conduction band is:

$$\varepsilon^{opt}(-1/0; e) = \Delta H_{form,TRAP}(E_C, 0) - \Delta H_{form,TRAP}(E_C, -1) = E_{TRAP, 0} - E_{TRAP,-1} + E_C$$

where, in the present study, $E_{TRAP,0}$ and $E_{TRAP,-1}$ are the neutral and charged total energies of the nanoparticle in the atomic structure where the electron is self-trapped and $E_C$ is the energy of the bottom of the conduction band (estimated by the computed vertical electron affinity, $E_C \approx -EA$). Analogously, the optical transition between the valence band and the hole trap state is:

$$\varepsilon^{opt}(+1/0; h) = \Delta H_{form,TRAP}(E_V, 0) - \Delta H_{form,TRAP}(E_V, +1) = E_{TRAP, 0} - E_{TRAP,+1} - E_V$$

where $E_V$ is the energy of the top of the valence band (estimated by the computed vertical ionization potential, $E_V \approx -IP$). The total energy of charged states are directly available in our calculations, because no periodic boundary conditions are applied.

## 3. Results and Discussion
## 3.1 Free/Trapped Excitons and Radiative Recombination
### 3.1.1 Experimental background

Photoluminescence (PL) measurements in anatase single crystals[34,35,36] show a broad band at ~2.3 eV with a full width of 0.6 eV, attributed to radiative recombination of a self-trapped exciton (STE) localized on a $TiO_6$ octahedron. A tail of the luminescence band reaching photon energies lower than 2.0 eV is often observed. The intensity of this emission decreases with increasing temperature: at RT it is almost quenched. The activation energy of this process is ~50 meV.[37]

A decomposition of the broad PL band into three Gaussian bands, centered at 1.95, 2.15 and 2.40 eV, was proposed.[38] The lowest energy emission would be probably closely related to the presence of oxygen vacancies since it is also present in the emission spectrum obtained at lower excitation energies. The other two emissions were assigned to the exciton (STE) recombination in bulk anatase.

At a sufficiently low temperature, an analogous STE emission feature is observed in the emission spectra of nanoparticles and nanocrystalline films with grain sizes in the range between 4 to 6 nm,[39] 9 to 27 nm[40] and 20 to 130 nm.[41] No significant variations with respect to bulk crystals were reported. The spectrum does not change with the nanocrystals size[39,40] or environment (water or thin film).[42] Only the dynamics of the photogenerated charges were found to be affected, because of a higher ratio between surface and bulk states for small nanocrystals, causing a faster transfer to surface traps. Thus, the trapping processes, which quenches the PL, occurs at lower temperature in ~5 nm nanoparticles than in bulk crystals.[34,38]

For very small nanoparticles (2-9 nm) a band-edge luminescence was reported at RT.[43,44,45,46] This emission was assigned to the fast radiative recombination of free (non trapped) excitons, which is not observed for bigger nanocrystals and bulk systems.[40] Finally, when present, interstitial $Ti^{3+}$ defects give rise to an additional PL peak at lower energies (~1.8 eV).

In the next paragraphs, the experimental findings reported above are discussed in the light of the computational results obtained in the present work for bulk, faceted and spherical nanoparticles of anatase $TiO_2$.

### 3.1.2 Bulk system

We first present the STE computed for the bulk anatase system, as described in detail in a previous work.[47] We have localized two possible trapped excitons which differ from the position of the hole trapping O atom ($O^-$ species) with respect to the electron trapping Ti atom ($Ti^{3+}$ species): this can be either in the equatorial ($Ti^{3+}$—$O_{eq}^-$) or in the axial ($Ti^{3+}$—$O_{ax}^-$) position of coordination in the $TiO_6$ octahedron. The two positions differ by a slightly longer Ti—O bond for the axial case which is confirmed experimentally by the XRD structure (exp. 1.98 Å;[48] calc. in this work 2.00 Å). The trapping energy of the two excitons differs by 0.1 eV in favour of the axial (-0.59 eV vs -0.49 eV, see **Table 1**). The computed PL emission energy associated to the decay of these two different trapped species is 1.99 eV ($Ti^{3+}$—$O_{ax}^-$) and 2.35 eV ($Ti^{3+}$—$O_{eq}^-$), respectively. The relative abundance of the equatorial species is twice that of the axial one. The emission energy values, together with the abundance ratio of 2:1, are in excellent agreement with the broad experimental feature[37] observed for anatase single crystals, just described in the previous section and centered at 2.3 eV.

### 3.1.3 Nanoparticles

3.1.3.a Faceted Nanoparticles

Excitons in nanoparticles may be very different from excitons in the bulk. We first present excitons in faceted nanocrystals, considering two models of different size, as shown in **Figure 1**: **NC$_S$** (489 atoms) and **NC$_L$** (798 atoms). The larger one presents an additional (101) layer on four of the eight (101) facets. This comparison will provide information on the size effect.

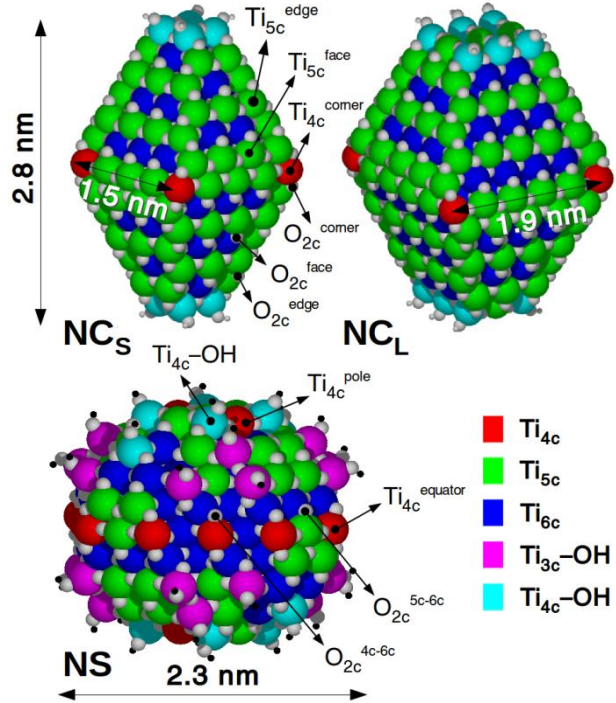

**Fig 1** Position of the Ti atoms with different coordination sphere within the nanoparticle models ($NC_S$ – $(TiO_2)_{159}$ • 4 $H_2O$; $NC_L$ – $(TiO_2)_{260}$ • 6 $H_2O$; $NS$ – $(TiO_2)_{223}$ • 18 $H_2O$) is visually shown by the color coding indicated on the right side. The O and Ti sites that were considered for charge trapping are labeled as in the tables. Relevant geometric parameter are also given.

**Table 1** Singlet-Triplet Vertical Excitation Energy ($S_0 \rightarrow T_1$)$_{vert}$, Trapping Energy ($\Delta E_{trap}$) of the Triplet Exciton and its calculated photoluminescence (PL) in the Axial and Equatorial configuration for bulk anatase as obtained with B3LYP functional, $NC_S$ and $NC_L$ nanocrystals and $NS$ and $NS_L$ nanospheres. All energies are in eV. Analogous HSE06 calculations are reported in **Table S1** in SI.

|  |  | Bulk | $NC_S$ | $NC_L$ | NS | $NS_L$ |
|---|---|---|---|---|---|---|
| ($S_0 \rightarrow T_1$)$_{vert}$ | | 3.91 | 3.95 | 3.95 | 3.88 | |
| $Ti^{3+} - O_{ax}^-$ | $\Delta E_{trap}$ | -0.59 | -0.90 | -0.80 | -0.65 | |
| | PL | 1.99 | 1.71 | 1.79 | 1.95 | 1.96 |
| $Ti^{3+} - O_{eq}^-$ | $\Delta E_{trap}$ | -0.49 | -0.55 | | | |
| | PL | 2.35 | 1.80 | | | |

Upon light irradiation of the small nanocrystals ($NC_S$), we expect the formation of a free exciton. This is computationally obtained by calculating the first excited triplet state in the Frank-Condon approximation (i.e. in the ground state geometry),[46] see vertical excitation in **Figure 2** and model **a** in **Figure 3**. Then, if we allow atomic relaxation in the excited state, we may observe the triplet exciton self-trapping (blue arrow in **Figure 2**). Indeed, we do, with two possible outcomes, in analogy to what observed for the bulk system: the trapping energy ($\Delta E_{trap}$ in **Table 1**) for the axial exciton (-0.90 eV), see model **b** in **Figure 3**, is larger than for the equatorial one (-0.55 eV). The charge delocalization in the free exciton compared to the charge localization in the self-trapped exciton is evident from the comparison of the spin density plots and of the histograms of the atomic spin for models **a** and **b**, respectively, as reported in **Figure 3**. Especially the histogram of model **b** proves the almost full localization of the unpaired electron and hole on two single Ti (black, 71 %) and O (red, 92 %) atoms, respectively (see **Table 2**).

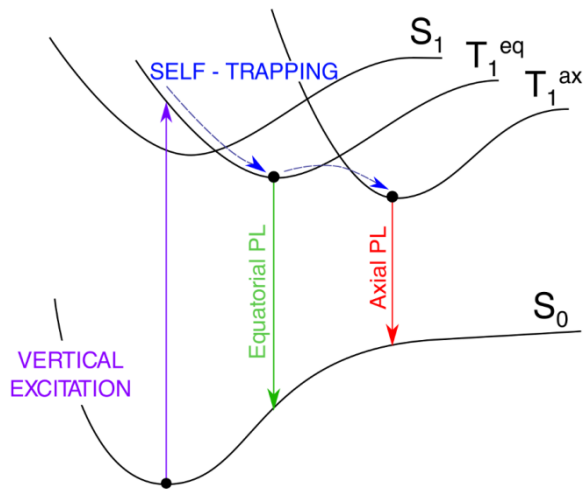

**Fig 2** Schematic representation of the processes involving the e$^-$–h$^+$ pair: the vertical excitation S$_0$ → T$_1$, the self-trapping relaxation in the bulk structure and the two different T$_1^{eq}$ → S$_0$ and T$_1^{ax}$ → S$_0$ emissions.

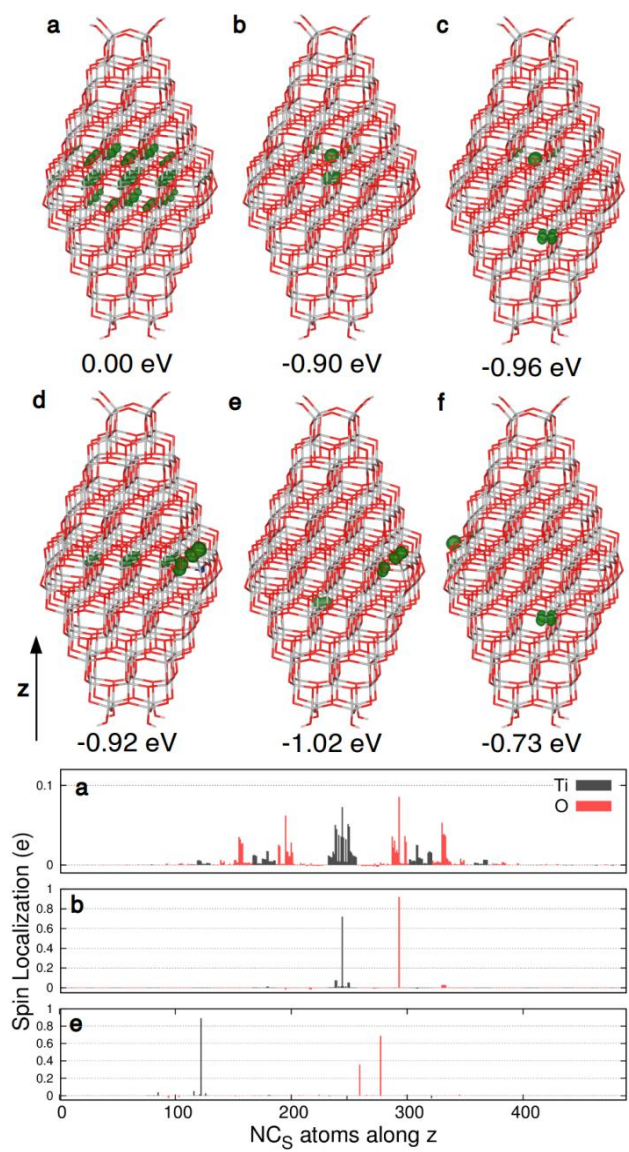

**Fig 3** Top panel: 3D spin density plots of **NC$_S$** nanocrystal, as obtained with the B3LYP functional, for the vertical triplet state (**a**), trapped triplet exciton (axial exciton in **Table 2**, **b**), triplet state with electron on the surface and hole in the core (**c**), triplet state with electron in the core and hole on the surface and completely separated charges state with the hole on O$_{2c}^{corner}$ (**e**) or O$_{2c}^{face}$ (**f**) and the electron on Ti$_{5c}^{face}$. The spin density isovalue is 0.01 e/Å$^3$ (0.002 e/Å$^3$ for the vertical triplet). Bottom panel: Histogram of the atomic spin on each O or Ti atoms for the a, b and c configurations. The x axis orders the atoms from bottom to top by their z coordinate.

The axial STE (model **b** in **Figure 3**) in the core of the small nanocrystal is a bound exciton, where the electron and the hole localize on a Ti—O bonded pair: Ti$^{3+}$—O$_{ax}^-$. If the electron and the hole succeed in separating within the nanocrystal by overcoming the small energy barrier (due to the exciton binding energy)[49] before recombination, the system may evolve in various possible final configurations, which are represented in **Figure 3** and detailed in **Table 2**: i) model **c**, where the electron has travelled to the surface to be trapped on an undercoordinated Ti$_{5c}$ atom on a (101) facet; ii) model **d**, where the hole has travelled to the surface to be trapped at an O$_{2c}$ site, leaving a delocalized electron in the core of the nanocrystal; iii) models **e** and **f**, where both electron and hole have reached the surface to be trapped at a Ti$_{5c}$ and an O$_{2c}$ site, respectively. It is interesting to note that an electron cannot be localized on a single Ti site in the nanocrystal, unless it is bound to the hole in the excitonic pair at the core of the nanocrystal or is trapped at a surface site.

Situations **c**, **d**, **e** and **f** exist if the time scale for migration and trapping at the surface is shorter than that of recombination. Distances in nanocrystals are short, the surface can be easily reached and therefore lower temperatures are required to quench the photoluminescence. However, on the contrary, if the STE of model **b** recombines before charge migration, the emission energy is estimated by current calculations to be 1.71 eV (see PL in **Table 1**). This value is rather low with respect to bulk value, probably due to the small size of the nanocrystal model used.

**Table 2** Trapping Energy ($\Delta E_{trap}$) of the triplet exciton at different sites with the charge localization (%electron or %hole) in **NC$_S$** and **NS$_S$** anatase nanoparticles, as obtained with the B3LYP functionals. No symmetry constrains are imposed to all the calculations. Energies are in eV. The sites nomenclature is defined graphically in **Figure 1.** Analogous HSE06 calculations are reported in **Table S2** in SI.

| Model | Position | $\Delta E_{trap}$ | %electron | %hole |
|---|---|---|---|---|
| | Electron/Hole pairs in **NC$_S$** | | | |
| b | Ti$_{6c}^{core}$ – O$_{3c}^{core}$ (Ti$^{3+}$ – O$_{ax}^-$) | -0.90 | 71% | 92% |
| c | Ti$_{5c}^{face}$ – O$_{3c}^{core}$ | -0.96 | 89% | 91% |
| d | Ti$_{6c}^{core}$ – O$_{2c}^{corner}$ | -0.92 | 13% | 63%/37% |
| e | Ti$_{5c}^{face}$ – O$_{2c}^{corner}$ | -1.02 | 89% | 63%/37% |
| f | Ti$_{5c}^{face}$ – O$_{2c}^{face}$ | -0.73 | 89% | 95% |
| | Electron/Hole pairs in **NC$_L$** | | | |
| | Ti$_{6c}^{core}$ – O$_{3c}^{core}$ (Ti$^{3+}$ – O$_{ax}^-$) | -0.80 | 74% | 91% |
| | Electron/Hole pairs in **NS** | | | |
| | Ti$_{6c}^{core}$ – O$_{3c}^{core}$ (Ti$^{3+}$ – O$_{ax}^-$) | -0.65 | 74% | 90% |
| | Ti$_{6c}^{subsurf}$ – O$_{2c}^{5c-6c}$ | -0.79 | 86% | 90% |

If we compare the results obtained for **NC$_S$** with those obtained for **NC$_L$** in **Table 2**, we may note some clear trends. The exciton trapping energy, for the axial model, decreases from -0.90 to -0.80 eV, while the PL emission increases from 1.71 eV to 1.79 eV (**Table 1**). This indicates that the bigger the model the larger the emission energy. Considering that the average radius of these two nanocrystals is 1.8 and 2.2 nm, one could extrapolate for nanocrystals of 4 nm size, reported in the experimental works, an emission energy close to 2 eV, similar to that for bulk anatase.

The % of electron-hole localization is quite high, except for the electron component in the **d** model. Here, the hole is far apart on the nanocrystal's corner and thus the electron prefers to delocalize on several Ti centers. On the contrary, in model **b**, where the hole is on the nearest-neighbor axial O atom, the exciton stabilization favors the electron localization on a single Ti site.

As a final remark, we wish to compare the B3LYP results with a set of HSE06 calculations (see **Table S1** in the SI). The latter functional provides similar energies for all the types of electron-hole pairs considered. Moreover, the HSE06 calculation for the model **b** shows a preference for the electron delocalization, even when the hole is localized on the nearest-neighbor axial O atom.

3.1.3.b Spherical Nanoparticles

We now present the excitons in spherical nanoparticles or nanospheres (**NS**), using as the reference model that reported in **Figure 1**. This type of nanoparticle is clearly more disordered than the faceted one, due to the presence of different classes of undercoordinated sites and due to the presence of a larger number of hydroxyl groups required to saturate the Ti atoms to a minimum four-fold coordination, which we have set for the chemical stability of the TiO$_2$ nanoparticles (see Computational Details).

The axial STE (Ti$^{3+}$—O$_{ax}^-$) in the core of spherical nanospheres (model **b** in **Figure 4**) is found to be less trapped than in the faceted models ($\Delta E_{trap}$ of -0.65 eV vs -0.90 or -0.80 eV for **NC$_S$** or **NC$_L$**, respectively in **Table 1**). The degree of localization of the electron and hole in the bound exciton is very high, similar to what observed for the corresponding model **b** in **NC$_S$**, if one compares the histograms of atomic spins in **Figure 3** and **Figure 4** or % data in **Table 2**.

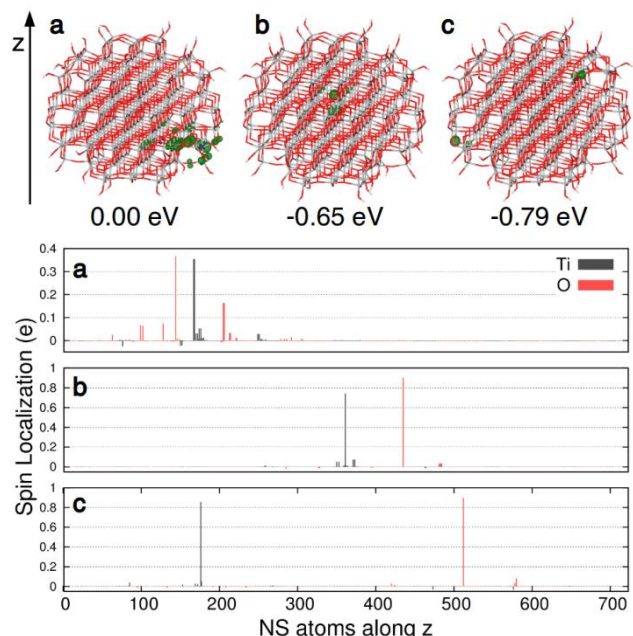

**Fig 4** Top panel: 3D spin density plots of **NS** nanosphere, as obtained with the B3LYP functional, for the vertical triplet state (**a**), trapped triplet exciton (**b**) and separated charges state (**c**) with the hole on $O_{2c}^{5c-6c}$ and the electron on $Ti_{6c}^{subsurf}$. Bottom panel: Histogram of the atomic spin on each O or Ti atoms for the a, b and c configurations. The x axis orders the atoms from bottom to top by their z coordinate.

If the electron and hole have enough energy to separate, they may travel to the surface, where they are trapped at atomic sites. In the case of the nanosphere we have just considered the most stable configurations for the electron and hole, respectively: the electron goes on a subsurface $Ti_{6c}$ site (see **Figure S1**), while the hole goes on a surface $O_{2c}$ site. The trapping energy in this configuration is larger and amounts to -0.79 eV. This result is a clear evidence that electrons and holes are preferentially trapped closer to or at the surface than in the core of the nanoparticle. The favourable energy gradient is the driving force of this natural process.

The core axial exciton in the nanosphere (**NS**) is characterized by a more similar environment to that in the bulk system than the analogous excitons in the small and large nanocrystals. This is because in the **NS** model the central Ti—O pair is surrounded by a larger number of next-neighbouring atoms in the x,y plane, i.e. the **NS** model has a larger equivalent diameter in the x,y plane than **NC_S** and **NC_L**. For this reason the computed PL emission energy (1.95 eV) associated to the STE is closer to that obtained for bulk (1.98 eV) than the one computed for the larger nanocrystal, **NC_L** (1.79 eV), in better agreement with experiments, showing that, at low temperature, PL spectra of nanoparticles do not differ much from PL spectra of single crystals. This is confirmed even by using a larger model (**NS_L**, described in the Computational details) with a PL value of 1.96 eV.

**3.2 Separate Carriers Trapping**

**3.2.1 Experimental background**

Two families of electron traps in anatase powder samples are commonly reported in the literature: short-living (< 50 ps) shallow traps,[3] in the range of 50–100 meV below the conduction band[50,51,52,53] and long-living deep traps, with trapping energies between 0.4 and 0.6 eV.[54,55,56] On the contrary, photogenerated holes are found to be deeply trapped in less than 200 fs,[49] so direct measurements of holes in shallow traps are not reported in literature. Carriers trapping can be probed by spectroscopic techniques, such as electron paramagnetic resonance (EPR) spectroscopy and transient absorption (TA) spectroscopy.

For example, integrating EPR intensities, Berger et al. determined that only one photoexcited $e^-$/$h^+$ pair is present per single UV-irradiated anatase nanoparticle (≈ 13 nm),[57] as a consequence of the coulombic repulsions between excess electrons. The majority of the photoexcited electrons (90%) are EPR silent, thus are delocalized. Only 10% of the electron are EPR active because they are localized at $Ti^{3+}$ sites with a de-trapping activation barrier estimated to be ≈ 26 meV. Photogenerated holes are found to be trapped at the surface of nanoparticles, as $O^-$ species, since they are completely scavenged by molecular oxygen.[58]

Regarding the source of the EPR signal, Giamello et al.[59] assigned the broad and unfeatured signal (g ≈ 1.93) to electrons trapped in the external layers of the nanocrystals. These trapping sites have slightly different local environment and therefore present slightly different EPR parameters. The other signal ($g_\parallel$ = 1.962, $g_\perp$ = 1.992) was assigned to $Ti^{3+}$ in regular bulk positions. According to the authors, the corresponding low hyperfine coupling with $^{17}O$ (< 2 MHz) is an evidence that this latter type of unpaired electron is quite delocalized over a discrete number of lattice sites (Bohr radius ≈ 15 Å). Results are found not to depend on the nanoparticles size in the range between 6 to 43 nm.

No differences in the orthorombic g-tensor of the hole signal were reported by Micic et al.[60] for anatase nanoparticles in the range between 2.5 and 30 nm in size. These authors concluded that $OH^\bullet$ is not the hole trapping species and that $O^{\bullet -}$ is bound to Ti with inequivalent crystal field splitting. More recently, Brezová et al.[61] measured two different axial g-tensor for photogenerated holes in nanoparticles, see Section 3.2.2.d: $O_{A1}^-$, which is dominant ($^{17}O$ hyperfine coupling constants are available) and $O_{A2}^-$, which was supposed to be close to more positively charged ions.

Transient absorption spectroscopy concerns the measurement of the differential absorption of irradiated and non-irradiated material samples. With this technique is possible to probe the transitions between electron trap levels and conduction bands, or between valence bands and hole trap levels. Generally, in these experiments, the estimated amount of $h^+$/$e^-$ pairs is less than 1–2 per metal oxide nanoparticle.

Recently, TA measurements in the mid-IR were used to determine the nature of shallow electron traps in 15–20 nm photoexcited anatase nanoparticles.[62] They were unambiguously attributed to polaronic states in the bulk of the nanoparticles and not to pre-existing defects. The radius of the polaron was estimated to be between 1 and 2 nm in an anatase single crystal.[63]

Regarding deep traps, the transient absorption (TA) spectrum in the visible to NIR range of anatase nanocrystalline films (grain size = 15–20 nm), mainly exposing (101) facets, have been decomposed by Yoshihara et al.[64]: the background spectrum was attributed to free electrons in the CB, whereas the broad absorption bands centered at 520 nm (2.38 eV) and 770 (1.61 eV) were assigned to trapped holes and electrons, respectively. These authors concluded that trapped carriers are at the surface, because they rapidly react with scavengers, whilst free electrons are in the bulk.

Similar measurements were taken by Shkrob et al. but for 4.6 nm aqueous anatase nanoparticles, with a nearly spherical shape, whose transient holes and electrons were detected at 650 nm (1.90 eV) and 900 nm (1.37 eV), respectively. Moreover, they observed that a small subset of holes are located in the interior of the nanoparticles since they cannot be scavenged by adsorbed species.[65]

Photoinduced electron traps in 20 nm faceted and 10 nm amorphous nanoparticles were recently studied by X-Ray absorption Spectroscopy with a 100 ps time resolution.[66] This study probed two types of $Ti^{3+}$ species under UV irradiation, which were assigned to non-symmetrically distorted hexacoordinated and pentacoordinated $Ti^{3+}$ sites. However, concomitant existence of shallower traps was not excluded.

**3.2.2 Electron Trapping**

3.2.2.a Faceted Nanoparticles

In this section, we investigate the electron trapping process in more detail by describing what happens when a single extra electron is present in the nanoparticles. In the case of a faceted model (**NC$_S$**), the extra free electron results to be largely delocalized on the majority of the Ti sites (see top left (**a**) of **Figure 5** and top panel (**a**) of **Figure S2**), in a quasi-band state. When the atomic positions are relaxed, the number of Ti sites where the electron is delocalized is reduced and only the central horizontal layers are involved (see top central (**b**) in **Figure 5** and central panel (**b**) of **Figure S2**). The energy gain associated with the atomic relaxation is of -0.21 eV. Finally, the local trapping may take place, which is not necessarily more advantageous because of the positive distortion energy contribution ($\Delta E_{dist}$). Various different sites have been considered in this study ranging from a $Ti_{4c}$ on a corner, $Ti_{5c}$ on a face, $Ti_{5c}$ on an edge, $Ti_{6c}$ subsurface (see **Table 3**, **Figure 1a** and **Figure S1** for nomenclature). Rather unexpectedly, edge and corner sites are not found to be involved in the trapping. This is because only the $Ti_{5c}$ on a (101) face and the $Ti_{6c}$ subsurface present a larger trapping energy ($\Delta E_{trap}$ = -0.28 eV) than the electron delocalized in the core layers of the nanoparticle (Core$^{deloc}$, -0.21 eV). The trapping energy value can be approximated more roughly in terms of the energy cost due to the distortion ($\Delta E_{dist}$) and the energy gain due to the occupation of a new Kohn-Sham state or SOMO ($\varepsilon_{SOMO}$). The best balance is observed (see **Table 3**) for the two aforementioned trapping sites. In the top right panel (**c**) of **Figure 5**, we only show the Ti$_{5c}$ on a (101) facet, because of the higher electron localization (89% vs 77%) with respect to the Ti$_{6c}$ subsurface, as described by the spin density plot, by the projected density of states in (**c**) and by the histogram of atomic spin in panel (**c**) of **Figure S2**.

For the larger faceted model (**NC$_L$**), trapping energies are systematically smaller or even positive due to a more stable and more delocalized reference states, however the same considerations described above may apply. Additionally, for this larger model, it has been possible to obtain electron localization on a single Ti ion in the deep core of the nanocrystal ((**h**) in **Figure S4** or Ti$_{6c}^{core}$ in **Table 3**). This solution, however, is isoenergetic with the free electron solution ((**f**) in **Figure S4**) and less stable than the one delocalized within the core ((**g**) in **Figure S4** or Core$^{deloc}$ in **Table 3**).

The energetics does not seem to significantly favour the single site trapping with respect to the electron relaxation in the nanoparticle core, which is quite interesting, given that a similar calculation for a bulk and slab model would favour the trapping site with respect to full delocalization in a crystal band state by -0.23 and -0.62, respectively.[46] We may conclude that the delocalization of the extra electron in the nanoparticle produces a quite stable large polaron, involving about 3-5 Ti atomic layer and extended from the center to the borders of the nanoparticle. This intermediate situation cannot be simply observed in periodic models of bulk and slabs because of the infinite character of bands.

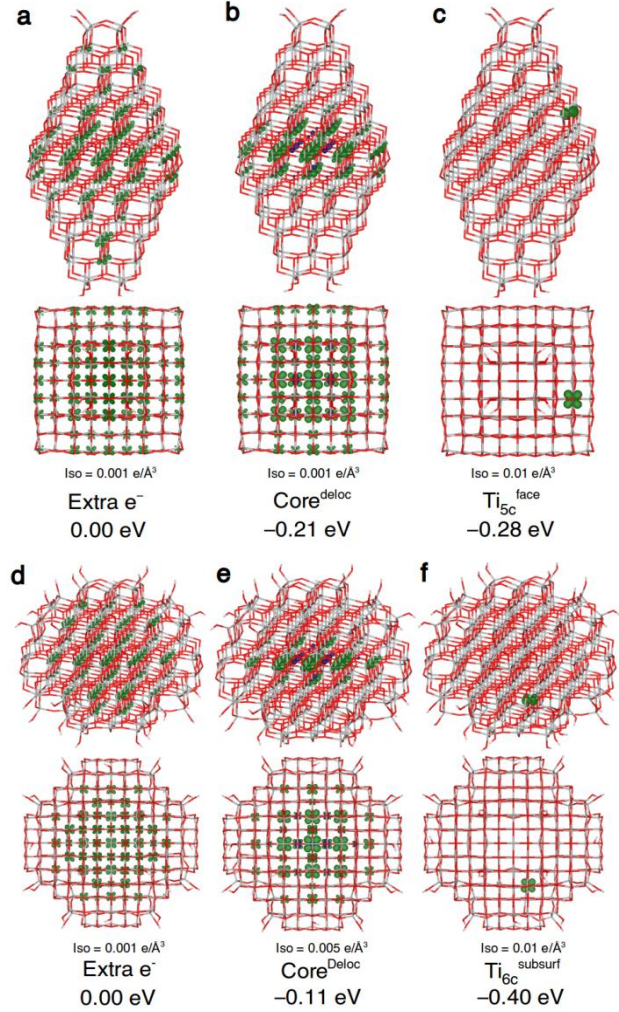

**Fig 5** Front and top view of the 3D plots of spin density of trapped electrons in **NC_S** nanocrystal (top panel) and in **NS** nanosphere (bottom panel), as obtained with the B3LYP functional. Below each structure the isovalue of corresponding 3D plot and the energy gain ($\Delta E_{trap}$) relative to the vertical addition of an excess charge are given. The sites nomenclature is defined graphically in **Figure 1** or in the text.

**Table 3** Trapping Energy ($\Delta E_{trap}$) for electrons at different sites for the three anatase nanoparticles with B3LYP functional. The reference zero for $\Delta E_{trap}$ is obtained by adding one electron with no atomic relaxation. The charge localization (%electron), distortion energy ($\Delta E_{dist}$) and eigenvalue of the unpaired electron state ($\varepsilon_{SOMO}$, respect to the bottom of the CB) are also given. Electron Affinity (EA) of one extra electron coming from the vacuum is reported. The potential versus the Standard Hydrogen Electrode ($U_{SHE}$) has been calculated using the approach in ref. 67 for B3LYP functional. No symmetry constrains are imposed in the calculations. Energies are in eV. The sites nomenclature is defined graphically in **Figure 1** and **Figure S1** or in the text. Analogous HSE06 calculations are reported in **Table S3** in SI.

| Position | $\Delta E_{trap}$ | %electron | $\Delta E_{dist}$ | $\varepsilon_{SOMO}$ | EA | $U_{SHE}$ (V) |
|---|---|---|---|---|---|---|
| Excess electron in **NC_S** | | | | | | |
| Vertical | | | | | 3.23 | -0.73 |
| $Ti_{4c}^{corner}$ | -0.10 | 90% | 1.26 | -1.07 | 3.33 | -0.63 |
| $Ti_{5c}^{face}$ | -0.28 | 89% | 1.10 | -1.20 | 3.51 | -0.45 |
| $Ti_{5c}^{edge}$ | -0.12 | 88% | 1.19 | -1.03 | 3.35 | -0.61 |
| $Ti_{6c}^{subsurf}$ | -0.28 | 77% | 0.64 | -0.88 | 3.51 | -0.45 |
| $Core^{deloc}$ | -0.21 | 11% | 0.03 | -0.24 | 3.44 | -0.52 |

|  | Excess electron in **NC$_L$** |  |  |  |  |  |
|---|---|---|---|---|---|---|
| Vertical |  |  |  |  | 3.42 |  |
| Ti$_{4c}^{corner}$ | 0.10 | 90% | 1.41 | -0.96 | 3.32 | -0.64 |
| Ti$_{5c}^{face}$ | -0.09 | 89% | 1.21 | -1.08 | 3.51 | -0.45 |
| Ti$_{6c}^{core}$ | 0.00 | 46% | 0.36 | -0.42 | 3.41 | -0.55 |
| Ti$_{6c}^{subsurf}$ | -0.09 | 78% | 0.77 | -0.80 | 3.51 | -0.45 |
| Core$^{deloc}$ | -0.04 | 4% | 0.12 | -0.11 | 3.46 | -0.50 |
|  | Excess electron in **NS** |  |  |  |  |  |
| Vertical |  |  |  |  | 3.15 | -0.81 |
| Ti$_{4c}^{equator}$ | -0.09 | 88% | 1.26 | -1.07 | 3.25 | -0.71 |
| Ti$_{4c}^{pole}$ |  | No trapping |  |  |  |  |
| Ti$_{4c}$–OH |  | No trapping |  |  |  |  |
| Ti$_{5c}$ |  | No trapping |  |  |  |  |
| Ti$_{6c}^{subsurf}$ | -0.40 | 85% | 0.85 | -1.11 | 3.55 | -0.41 |
| Ti$_{6c}^{core}$ | -0.13 | 62% | 0.47 | -0.61 | 3.28 | -0.68 |
| Core$^{deloc}$ | -0.11 | 19% | 0.17 | -0.26 | 3.26 | -0.70 |
|  | Excess electron in **Bulk Anatase**[46] |  |  |  |  |  |
| Bulk | -0.23 | 80% | 0.42 | -0.82 |  |  |

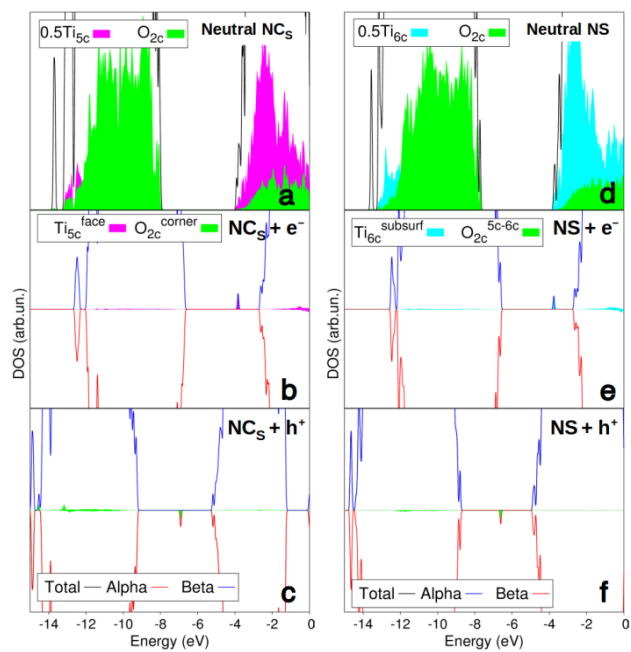

**Fig 6** Simulated total (DOS) and projected (PDOS) density of states on different O and Ti atoms, as calculated with the B3LYP functional, of: (**a**) the neutral **NC$_S$**, (**b**) **NC$_S$** with an additional electron and (**c**) **NC$_S$** with an additional hole in their best trapping site; (**d**) the neutral **NS**, (**e**) **NS** with an additional electron and (**f**) **NS** with an additional hole in their best trapping site. A 0.001 eV Gaussian broadening was used. The zero energy is set to the vacuum level. The sites nomenclature is defined graphically in **Figure 1** and **Figure S1** or in the text.

3.2.2.b Spherical Nanoparticles

In the case of a spherical nanoparticle (**NS**), the extra electron is found to be fully delocalized on all the Ti centers, except those in the most external atomic shell. This is different from what observed for **NC$_S$** and **NC$_L$** and is clearly described by the spin density plot in the top panel (**d**) of **Figure 5** and by the histogram of atomic density in the top panel (**d**) of **Figure S2**. When atomic relaxation is allowed in the presence of the extra electron, a clear effect can be observed: the spin density localizes on three Ti atomic layers, with an associated energy gain of -0.11 eV (see panel (**e**) in **Figure 5** but also the histogram of atomic spin in panel (**e**) of **Figure S2**). Finally, electron localization on a single atomic site has also been investigated, as reported in **Table 3** for **NS**. A number of $Ti_{4c}$ and $Ti_{5c}$ sites were considered but, unexpectedly, no trapping was observed, except for a very tiny trapping energy (-0.09 eV) in the case of a four-fold Ti at the equator of the nanosphere (see $Ti_{4c}^{equator}$ in **Figure 1**). On the contrary, the most stable trapping site is determined to be a subsurface $Ti_{6c}$ (see **Figure S1**) with $\Delta E_{trap}$ of -0.40 eV. This site is describe in panel (**f**) of **Figure 5**. The high electron localization by 85% at this trap is confirmed also by the PDOS reported in **Figure 6e** and by the high atomic spin in panel (**f**) of **Figure S2**. No other trapping site (**Figure 1c**) seems to be competitive with this one (**Table 3**). Noteworthy is that, differently to what observed for faceted nanoparticles, here the electron delocalization at the core of the nanoparticle seems to be less effective if compared to full localization: -0.11 vs -0.40 eV.

If we focus the attention on EA in **Table 3**, we may directly compare electron trapping sites in different nanoparticles (**NC_S**, **NC_L** and **NS**), since for EA there is a common reference, which is the vacuum. We observe that the EA values are in a rather small range of energies: from 3.25 to 3.55 eV, independently of the size and shape of the nanoparticle, which suggests similar trapping abilities of the three nanoparticles considered.

3.2.2.c Trapping centers

In the top panel of **Figure 7**, the coordination spheres of a six-fold Ti atom in the anatase relaxed bulk (**Figure 7a**), of a five-fold Ti atom on the anatase (101) relaxed surface (**Figure 7b**) and of a four-fold Ti atom at a corner of a nanocrystal (**Figure 7c**) are represented through a ball-and-stick model, respectively. Bond distances and angles are reported as reference values to be compared with the analogous sites in the nanoparticle once it has trapped one electron. For both the faceted and spherical nanoparticles, we observe that the electron trapping process induces a general elongation of the bond distances, especially of those bonds in the plane of the Ti d orbital hosting the extra electron. On the other hand, bond angles do not change much with respect to the original site.

We also show another potential trapping site, a four-fold Ti ion (right side of **Figure 7**) that is present on both **NC_S** and **NS**. Here, the angle distortion is larger, while the bond elongation is less pronounced, both effects due to the absence of a second coordinating (and repulsive) O atom.

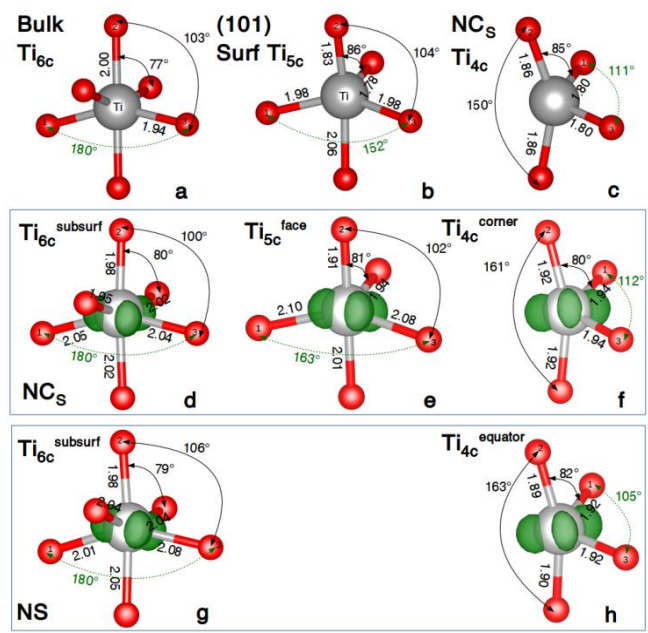

**Fig 7** Ball-and-stick representation and spin densities (Isovalue = 0.03 e/A$^3$) of the main trapping sites of an excess electron in **NC$_S$** (**d**, **e**, **f**) and **NS** (**g** and **h**), together with relevant bond lenghts and angles. As reference, the geometrical parameters for a **a**) Ti$_{6c}$ of an anatase bulk; **b**) Ti$_{5c}$ of an (101) anatase slab and **c**) Ti$_{4c}$ at the corner of **NC$_S$** are shown in the top panel. The dashed green line indicates the dihedral between the planes O3-Ti-O2 and O2-Ti-O1. The sites nomenclature is defined graphically in **Figure 1** and **Figure S1** or in the text.

3.2.2.d Comparison with experiments

Independently of the nanoparticle shape, we could find both shallow (delocalized) and deep (localized) self-trapping states for electrons, involving d$_{xy}$ states of Ti ions in the lattice. One should expect to observe high values of electron paramagnetic (EPR) hyperfine coupling constants with the next-neighboring $^{17}$O nuclei for the former and low values for the latter. As a matter of fact, we notice that when the electron is deeply trapped, and thus localized, the calculated hyperfine coupling constants with $^{17}$O are very similar to those computed for a model Ti$^{3+}$(H$_2$O)$_6$,[68] while there is a significant decrease of the isotropic coupling constant when the electron is delocalized over many Ti sites, in better agreement with EPR experimental values for anatase nanoparticles (a$_{iso}$ < 2 in **Table 4**).[59]

**Table 4** Calculated (B3LYP) and experimental hyperfine Fermi contact term with $^{17}$O (a$_{iso}$, in MHz) for Ti$^{3+}$(H$_2$O)$_6$ ion and for localized/delocalized models of one excess electron in anatase faceted and spherical nanoparticles. The degree of localization is expressed in % of an electron.

|  | System | Site | Localization | a$_{iso}$ | Reference |
|---|---|---|---|---|---|
| | Ti$^{3+}$(H$_2$O)$_6$ | - | 100% | 6.7 | 68 |
| **Calculated** | **NC$_L$** | Ti$_{6c}^{core}$ | 46% | 6.3 | |
| | | Core$^{deloc}$ | 4% | 1.4 | |
| | **NS** | Ti$_{6c}^{core}$ | 62% | 6.7 | |
| | | Core$^{deloc}$ | 19% | 3.9 | |
| **Experimental** | (Ti$^{3+}$)$_{aq}$ | - | Localized | 7.4 | 68 |
| | Anatase NP | - | Delocalized | <2 | 59 |

The calculated trapping energy, relative to a free electron in CB, is in good agreement with the experimental values: shallow traps are in the range between 0.04 to 0.20 eV,[50,51,52,53] whereas deep traps can reach 0.40 eV.[54,55,56] The best trapping sites are surface five-fold coordinated and subsurface six-fold coordinated Ti atoms, in both faceted and spherical nanoparticles, as evidenced by XAS spectroscopy.[66]

Finally, we compare the energies associated to the electronic transitions, as computed with the transition level approach (see Computational Details in Section 2), with the experimental absorption energies. The optical transitions for electron and hole traps in the core and on the surface of the nanoparticles are reported in **Table 5**.

**Table 5** Calculated (B3LYP) optical transitions between the electron trap levels and the CBM, $\varepsilon^{opt}(-1/0; +e)$, and the VBM and the hole trap levels, $\varepsilon^{opt}(+1/0; +h)$, for the three anatase nanoparticles under investigation. Results for the best trapping electron/hole trapping sites are highlighted in bold. The sites nomenclature is defined graphically in **Figure 1** and **Figure S1** or in the text.

| Optical Transition | Location | Trapping Site | $NC_S$ | $NC_L$ | NS |
|---|---|---|---|---|---|
| from $e^-$ trap to CBM $\varepsilon^{opt}(-1/0; +e)$ | Core | $e^-$ shallow | 0.24 | 0.16 | 0.28 |
| | | $Ti_{6c}^{core}$ | - | 0.37 | 0.60 |
| | Surface | $Ti_{6c}^{subsurf}$ | 0.91 | 0.87 | **1.25** |
| | | $Ti_{5c}^{face}$ | **1.38** | **1.30** | - |
| | Experimental | | | 1.61[a] | 1.37[b] |
| from VBM to $h^+$ trap $\varepsilon^{opt}(+1/0; +h)$ | Core | $O_{3c}^{core\_ax}$ | 2.44 | 2.32 | 2.07 |
| | Surface | $O_{2c}^{4c/5c}$ | **2.65** | **2.69** | 2.38 |
| | Experimental | | | 2.38[a] | 1.90[b] |

[a] From ref. 64
[b] From ref. 65

As mentioned above, the $Ti_{5c}^{3+}$ is found to be the best trapping site for faceted nanoparticles. The calculated transitions from this level to the CB minimum are 1.38 and 1.30 eV for **$NC_S$** and **$NC_L$**, respectively, which are a slightly smaller than the experimental value of 1.61 eV measured by transient absorption spectroscopy (TA) for films of anatase nanoparticles exposing mainly (101) facets.[64] This may be due to the fact that the electron is initially trapped in a $d_{xy}$ state ($t_{2g}$ state) and the final states which have the correct symmetry for an allowed transition ($e_g$) are positioned (see the PDOS in **Figure S7**) at least 0.5 eV above the CB minimum.

In the case of spherical nanoparticles, the best trapping site for an electron is the $Ti_{6c}^{3+}$ (see **Figure S1**). The calculated transition from this trap level to the CBM is 1.25 eV, in good agreement with the experimental value of 1.37 eV, observed by TA spectroscopy of aqueous spherical nanoparticles.[65]

Noteworthy is that the electronic transitions from species in the core or in subsurface layers are characterized by rather low excitation energies, especially in the case of a trapped electron in the core ($Ti_{6c}^{core}$).

**3.2.3 Hole Trapping**

3.2.3.a Faceted Nanoparticles

The presence of an electron hole in a faceted nanoparticle ($NC_S$) involves two full O layers at the center of the model (see panel (**a**) of **Figure 8**). After atomic relaxation, however, the delocalization is totally quenched and the hole results to be fully localized on a single three-fold coordinated O atom in the core of the nanocrystal in axial position with respect to the central Ti ion, with a considerable associated energy gain of -0.97 eV ($O_{3c}^{core\_ax}$ in panel (**b**) of **Figure 8**). If the hole has enough energy to overcome the barrier for diffusion (estimated to be of about 0.5 eV),[69] it may reach the corner of the nanocrystal ($O_{2c}^{corner}$, with a further energy gain of -0.11 eV, $\Delta E_{trap}$ = -1.08 eV). Two O atoms share the spin density of this corner site: an $O_{2c}$ and an $O_{3c}$, as shown in panel (**c**) of **Figure 8** and in the PDOS of **Figure 6**. Other possible, but less effective, trapping sites have been considered, which are documented in **Table 6** and represented in **Figure 1a** and **Figure S5**, together with the distortion energies and the Kohn-Sham eigenvalue of the unoccupied eigenstate describing the electron hole. Analogous results are obtained for the larger nanocrystal ($NC_L$), as reported in **Table 6** and, with the only difference of slightly smaller, but consistently shifted, trapping energies.

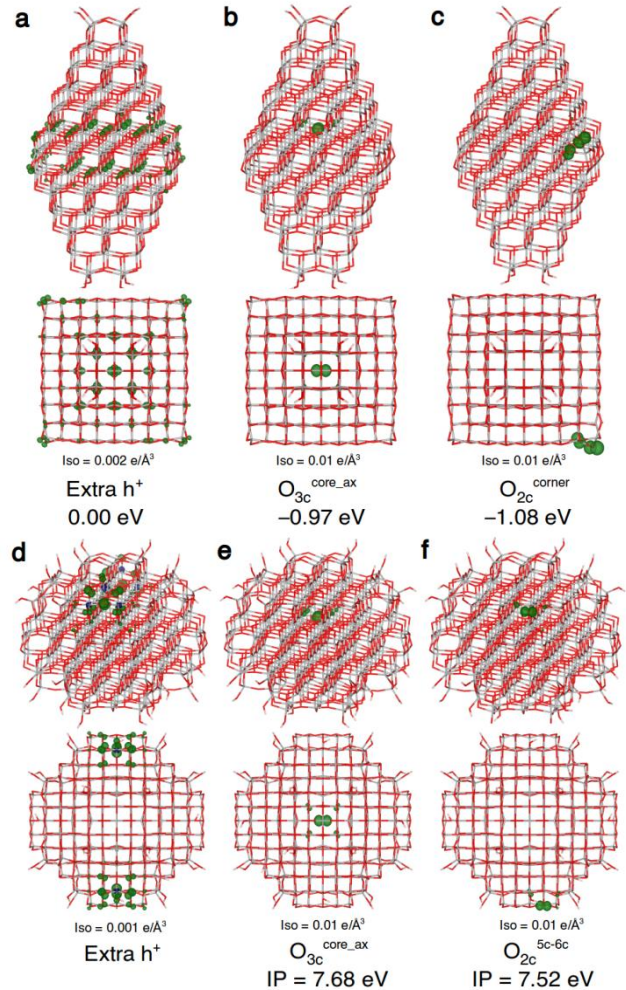

**Fig 8** Front and top views of the spin density 3D plots of an excess hole in **NC_S** nanocrystal (top panel) and **NS** nanosphere (bottom panel) for the main trapping sites, as obtained with B3LYP functional. Below each structure the isovalue of corresponding 3D plot and the energy gain ($\Delta E_{trap}$ in **Table 6**) relative to the vertical addition of an excess charge are given. The sites nomenclature is defined graphically in **Figure 1** or in the text.

**Table 6** Trapping Energy ($\Delta E_{trap}$) for an excess hole at different sites for the three anatase nanoparticles with B3LYP functional. The reference zero for $\Delta E_{trap}$ is obtained by removing one electron with no atomic relaxation. Their charge localization, distortion energy ($\Delta E_{dist}$) and eigenvalue of the hole state ($\varepsilon_{HOLE}$, respect to the top of the VB) are also given. Ionization Potential (IP) of the most external, less bound, electron to the vacuum level is reported. No symmetry constrains are imposed in all the calculations. Energies are in eV. The sites nomenclature is defined graphically in **Figure 1** or in the text. Analogous HSE06 calculations are reported in **Table S4** in SI.

| Position | $\Delta E_{trap}$ | Localization | $\Delta E_{dist}$ | $\varepsilon_{hole}$ | IP |
|---|---|---|---|---|---|
| Hole addition in **NC_S** | | | | | |
| Vertical | | | | | 8.70 |
| $O_{2c}^{face}$ | -0.77 | 95% | 1.75 | 2.32 | 7.93 |
| $O_{2c}^{corner}$ | -1.08 | 66%/33% | 1.67 | 2.35 | 7.62 |
| $O_{2c}^{edge}$ | -0.54 | 71%/29% | 1.82 | 2.06 | 8.16 |
| $O_{3c}^{core\_ax}$ | -0.97 | 91% | 1.47 | 2.49 | 7.73 |
| $O_{3c}^{core\_eq}$ | -0.59 | 83% | 1.30 | 1.88 | 8.11 |
| Hole addition in **NC_S**–OH | | | | | |
| $Ti_{5c}$–OH | | 70%/30% | 2.21 | 2.75 | 7.72 |
| Hole addition in **NC_L** | | | | | |
| Vertical | | | | | 8.64 |
| $O_{2c}^{corner}$ | -1.01 | 66%/33% | 1.78 | 2.35 | 7.63 |
| $O_{3c}^{core\_ax}$ | -0.84 | 90% | 1.48 | 2.32 | 7.80 |
| Hole addition in **NS** | | | | | |
| Vertical | | | | | 8.34 |
| $O_{2c}^{4c-6c}$ | | 89%/11% | 1.78 | 1.96 | 7.74 |
| $O_{2c}^{5c-6c}$ | | 92%/8% | 1.77 | 2.18 | 7.52 |
| $O_{3c}^{core\_ax}$ | | 90% | 1.41 | 1.98 | 7.68 |
| $Ti_{3c}$–OH | | No trapping | | | |
| $Ti_{4c}$–OH | | No trapping | | | |
| Hole addition in **NS–OH** | | | | | |
| $Ti_{5c}$–OH | | 85%/15% | 2.41 | 2.02 | 7.81 |

3.2.3.b Spherical Nanoparticles

In the case of spherical nanoparticles, given the lower degree of order and crystallinity, the removal of one electron, without any atomic relaxation (vertical ionization) involves only some portions of the nanoparticle, as evident by the spin density plot in panel (**d**) of **Figure 8**. This cannot be considered a delocalized band-like state and, therefore, should not be used as a reference to compute the trapping energies of the various hole trapping sites ($\Delta E_{trap}$), as was done for faceted nanoparticles. Thus, in the case of nanospheres, we will not compare trapping energies but adiabatic ionization potentials (IPs), which have a common reference that is the vacuum level. The smaller the adiabatic IP, the larger is the trapping energy of the site. Within the bulk of the nanosphere, the hole is found to be almost fully localize (90%) in the central three-fold coordinated O atom ($O_{3c}^{core\_ax}$). The adiabatic IP for this trapping site is 7.68 eV. If the hole has sufficient energy to diffuse, it may reach the surface. On the nanosphere surface, several types of O species exists (**Figure 1c**). The most stable trapping site is found to be a two-fold O atom bridging a $Ti_{5c}$ and a $Ti_{6c}$ atom (IP = 7.52 eV), see $O_{2c}^{5c-6c}$ in panel (**f**) in **Figure 8**. The distortion energy ($\Delta E_{dist}$) for this trapping site is larger than at the core of the nanoparticle (1.77 vs 1.41 eV), however, also the energy gain in terms of the energy level of the hole state ($\varepsilon_{hole}$) with respect to the top of the valence band is larger (2.18 vs 1.98 eV), see **Table 4** and **Figure 6f**. A second type of $O_{2c}$ is between a $Ti_{4c}$ and a $Ti_{6c}$, which is also a good trapping site but less efficient than the previous one (IP = 7.74 eV).

If we focus the attention on IPs in **Table 6**, we may directly compare electron trapping sites in different nanoparticles (**NC$_S$**, **NC$_L$** and **NS**), since for IPs there is a common reference, which is the vacuum. We observe that the IP values, considering all nanoparticles of various size and shape, are in the energy range between 7.52 and 8.16 eV. The nature of the best trapping sites on a nanocrystal and on a nanosphere is very different: for **NC$_S$** it is an $O_{2c}$ bound to a $Ti_{4c}$ at a corner site ($O_{2c}^{corner}$), while for **NS** it is an $O_{2c}$ bound to $Ti_{5c}/Ti_{6c}$ ions ($O_{2c}^{5c-6c}$).

3.2.3.c Trapping centers

In the top panel of **Figure 9** bond distances and angles for the three-fold coordinated O atom in the anatase relaxed bulk (**Figure 9a**), for the two-fold coordinated O atom on the anatase (101) regular relaxed surface (**Figure 9b**) and for a two-fold coordinated O atom at a corner of a nanocrystal (**Figure 9c**) are reported, respectively. Below the hole trapping sites are represented for both nanocrystals and nanospheres. The highest distortion is observed for the $O_{2c}$ at a corner site in the nanocrystals. Ti—O bond are elongated, while the O---O distance is shortened (from 2.45 to 2.22 Å). For all the other $O_{2c}$ species, only Ti—O bonds are elongated. In the case of $O_{3c}$, upon hole trapping, Ti—O bonds are longer (e.g. Ti—$O_{ax}$ by 0.3 Å) and the TiOTi angle decreases (from 103° to 98°).

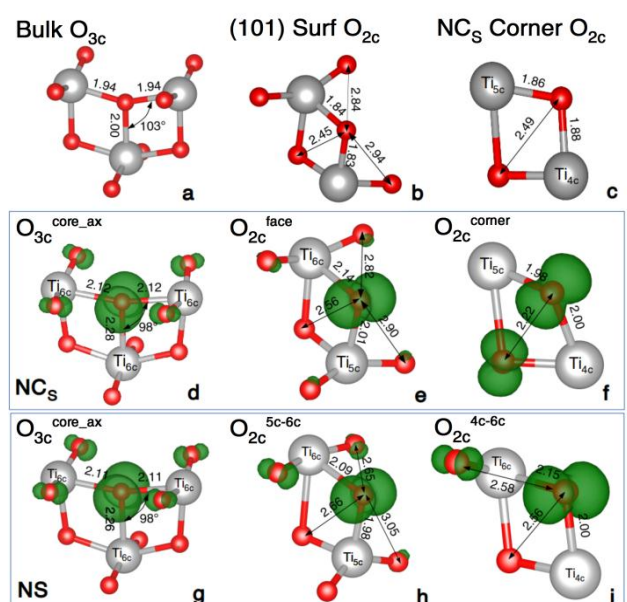

**Fig 9** Ball-and-stick representation and spin densities (Isovalue = 0.01 e/A$^3$) of the main trapping sites for an excess hole in **NC$_S$** (**d**, **e**, **f**) and **NS** (**g**, **h**, **i**), together with relevant bond lenghts and angles, as obtained with the B3LYP functional. As reference, the geometrical parameters for an **a**) $O_{3c}$ of an anatase bulk; **b**) $O_{2c}$ of an (101) anatase slab and **c**) $O_{2c}$ at the corner of **NC$_S$** are shown in the top panel. The sites nomenclature is defined graphically in **Figure 1** or in the text.

3.2.3.d Hydroxyl groups

Finally, another very important species, which is probably very common in aqueous solution and which might be a stable trapping site, is the OH group. We have found that OH bonded to five-coordinated Ti atoms are able to trap efficiently an excess hole. OH groups bonded to less coordinated Ti atoms, in particular $Ti_{3c}$−OH and $Ti_{4c}$−OH sites in **Figure 1**, could not trap an electron hole, despite the many attempts. We believe that the reason is that OH bonded to a fully coordinated Ti ion is more electron rich than an OH bonded to a lower coordinated Ti, therefore the former is a good hole trap, while the latter is a bad hole trap.

Since in the models of both faceted and spherical nanoparticle used there are no OH groups bound to $Ti_{5c}$ sites, the addition of a dissociated water molecule is necessary, resulting in a $Ti_{5c}$–OH and $O_{2c}$H (see models in **Figure 10**). To avoid any spurious interaction we have put the two fragments on opposite sides of the nanocrystal or of the nanosphere.

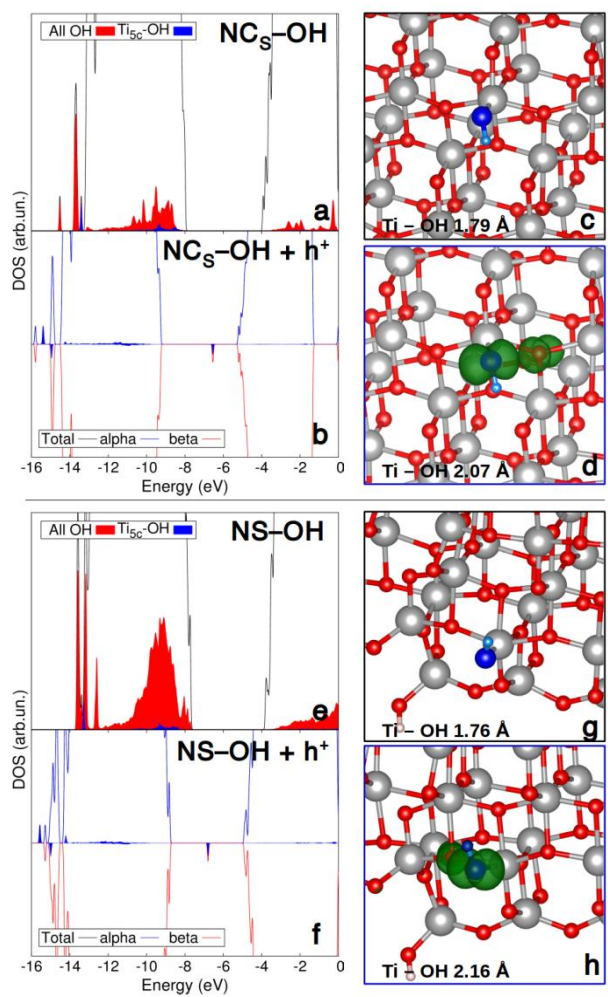

**Fig 10** Simulated total (DOS) and projected (PDOS) density of states on all the OH groups of the nanoparticle and on the specific trapping site, as calculated with the B3LYP functional, for the neutral **NC_S** and **NS** nanoparticles (**a** and **e** respectively) and for the same nanoparticles with an additional hole on the OH group (**b** and **f**). A 0.001 eV Gaussian broadening was used. The zero energy is set to the vacuum level. Ball-and-stick representation of a magnified portion of the neutral $TiO_2$ anatase nanocrystal **NC_S** (**c**) and nanosphere **NS** (**g**) containing the extra OH bound to $Ti_{5c}$, highlighted in blue. Spin density 3D plots (isovalue = 0.005 e/A$^3$) of the OH group when an electron is removed and the hole localizes on the hydroxyl group (**d,h**) are also given.

In the case of the nanocrystal (**NC_S**), through the comparison of IPs in **Table 6**, we observe that the additional OH group on a $Ti_{5c}$ (**NC_S-OH**) presents a higher trapping ability (lower IP, 7.72 vs 7.93 eV) than that of an $O_{2c}$ on a (101) facet ($O_{2c}^{face}$) but lower trapping ability (higher IP, 7.72 vs 7.62) than that of the $O_{2c}$ at a corner site ($O_{2c}^{corner}$). The hole is mainly localized (70%) in a p state of the O atom of the OH group and partially localized (30%) on a neighbour $O_{3c}$ atom, as clearly shown by the spin plot (**c**) in **Figure 10**.

In the case of the nanosphere, the trapping ability of the additional OH species (**NS-OH**) is found to be smaller than those of the bulk and surface O ions, since the IP is computed to be 7.81 eV, which is 0.29 eV higher than the value for the best trapping site ($O_{2c}^{5c-6c}$). The hole is highly localized in a p state of the O atom in the OH group (0.85) with some little delocalization on a nearby $O_{3c}$ (0.15), as clearly shown by the spin plot (**h**) in **Figure 10**.

Upon trapping, the Ti-O bond is found to elongate (from 1.79 to 2.07 Å in **NC$_S$-OH** and from 1.79 to 2.16 Å in **NS-OH**), whereas the OH bond is almost unaffected (from 0.974 to 0.991 Å in **NC$_S$-OH** and from 0.973 to 0.986 Å in **NS-OH**). From the total and projected DOS of the neutral nanoparticles (panels (**a**) and (**e**) of **Figure 10**), we may notice that the OH states (blue projection) are not at the top of the O 2p valence band. Once the electron is remove and the hole is formed (panels (**b**) and (**f**) of **Figure 10**) a new empty state is formed in the middle of the band gap. From the projected DOS it is evident that the hole state is almost fully localized on the OH group under investigation (blue projections).

If we compare the IP values for **NC$_S$-OH** and **NS-OH**, 7.72 vs 7.81 eV, we may conclude that the values are very close. Since the chemical nature of the two $Ti_{5c}$-OH species in the differently shaped nanoparticles is analogous, only small differences in the trapping abilities were expected.

3.2.3.e Comparison with experiments

In the case of holes, despite various attempts, we could localize only deep traps. The self-trapped hole at a bridging oxygen bound to a $Ti_{5c}$ and a $Ti_{6c}$ atom on the surface of both nanocrystals and nanospheres is the most stable. The calculated vertical electronic transition at this defect site is 2.52 eV for faceted nanoparticles, in satisfactory agreement with the reported TA value of 2.38 eV.[64] For spherical nanoparticles, we calculated a vertical transition of 2.59 eV, which is quite different from the reported experimental value of 1.9 eV by transient absorption spectroscopy.[65] The discrepancy is probably due to the fact that these experiments were performed on aqueous nanoparticles. Some preliminary data from our group show that the presence of dissociated or undissociated water molecules on the nanoparticle surface may affect the trapping processes, in agreement with a recent experimental work.[70] However, this issue is beyond the focus of the present work and will be the subject of a future study.

Nonetheless, the g- and A-tensor calculated for the self-trapped hole species (see values in **Table 7** and **Table S5**) are in exceptional quantitative agreement with those reported in the experimental literature.[59,60,61] This observation allows the conclusion that the computational models provide the correct degree of localization of unpaired electrons. The g-tensors have been calculated with the Gaussian09 code, using a smaller cluster (in **Figure S6**) cut from the faceted nanocrystal. Note that atomic relaxations of these self-trapped holes in the smaller cluster model were preliminarily performed with the CRYSTAL14 code, to be consistent with all the other calculations in this study.

**Table 7** g-tensor and hyperfine coupling constants (hpcc, in G) tensor with $^{17}$O, decomposed as the Fermi contact term ($a_{iso}$) and the dipolar tensor (B) for an excess hole at the $O_{2c}^{face}$ site in the $Ti_{29}O_{58}$ cluster, as calculated at the B3LYP/EPR-II level. Corresponding experimental data from literature for anatase nanoparticles are also given. The sites nomenclature is defined graphically in **Figure 1** or in the text.

| Position | $g_{xx}$ | $g_{yy}$ | $g_{zz}$ | $a_{iso}$ | $B_{aa}$ | $B_{bb}$ | $B_{cc}$ |
|---|---|---|---|---|---|---|---|
| $O_{3c}^{core}$ | 2.004 | 2.015 | 2.019 | -12 | 43 | 42 | -85 |
| $O_{2c}^{face}$ | 2.006 | 2.017 | 2.020 | -12 | 43 | 43 | -86 |
| Exp. $O_{A1}^{-}$ [61] | 2.004 | 2.019 | 2.019 | ±24 | 41 | 41 | -82 |
| Exp. $O_{A2}^{-}$ [61] | 2.004 | 2.014 | 2.014 | | | | |
| Exp. $O^{-}$ [59] | 2.003 | 2.015 | 2.027 | | | | |
| Exp. $O^{-}$ [60] | 2.007 | 2.014 | 2.024 | ±22 | 41 | 41 | -82 |

### 3.2.4 Distortions due to the Nanosized Confinement and Trapping

In this section we analyze how the distortions due to the particles nanosize may affect the charge trapping process. As already mentioned, in small nanocrystals the close presence of surfaces, edges and corners affects the Ti–O and Ti–Ti bond lengths with respect to the bulk values, even in the core of the model. We shall recall that, with the present computational setup, the Ti–O equatorial and axial bond lengths for the relaxed bulk anatase $TiO_2$ are 1.94 and 2.00 Å, respectively, and the Ti–Ti equatorial distance is 3.79 Å.

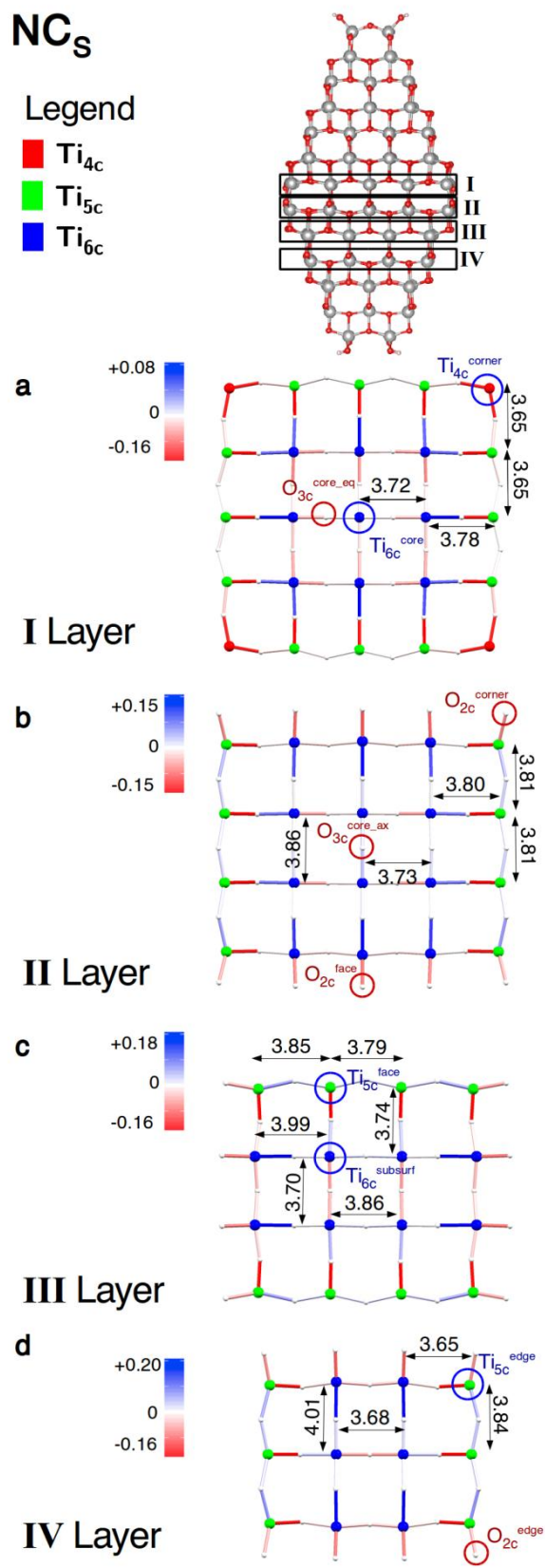

**Fig 11** Visual representation of the elongation/compression of the bonds within four layers of the **NC$_S$** nanocrystal from the B3LYP bulk values for equatorial and axial distances respectively. The elongation is given in shade of blue, the compression in shade of red, as indicated by the color scale close to each panel. Ti atoms are colored according to their coordination patterns, following the color code on the upper left. O atoms are represented as small white balls. Ions (labeled as in **Figure 1** and **Figure S1**) where the charge is highly localized localized are circled in blue (for electrons) or red (for holes). Distances between Ti atoms are in Å.

In **Figure 11**, equatorial distances are reported in detail. Those Ti—O distances are relevant for the electron trapping process since, as we have shown above, it involves the $d_{xy}$ orbitals. Several XY layers of **NC$_S$** are considered.

Looking at the first (central) layer (**Figure 11a**), we can clearly note that a Ti$_{5c}$ atom at the border causes a contraction of the Ti—O bond underneath (red), which is balanced by an elongation (blue) of the O—Ti$_{6c}$ in the next inner shell. Following the Ti–O–Ti line towards the core, the next Ti$_{6c}$—O distance is compressed, while the last O–Ti$_{6c}^{core}$ is almost unchanged. The overall result is a compression of the innermost Ti–Ti distance to 3.72 Å, significantly shorter than that in the bulk (3.79 Å). A shorter Ti–Ti distance limits the possibility of the Ti–O bonds elongation, which is a necessary structural modification to efficiently trap an excess charge on Ti (electron) or on O (hole) atoms. Moreover, as previously shown, the excess charge tends to localize on a Ti $d_{xy}$ or on an O $p_x$ (or $p_y$) orbital, so it is very sensitive to equatorial distances. Thus, we argue that this short Ti–Ti distance is the reason why no charge localization could be observed on the Ti$_{6c}^{core}$ site of **NC$_S$** and the trapping energy of the hole in O$_{3c}^{core\_eq}$ was lower than that of the bulk anatase.

At the corner of the first layer, in correspondence to the Ti$_{4c}$ trapping site, the highest compression of the Ti–O bond (–0.16 Å) and Ti–Ti distance (to 3.65 Å) is observed. These values can explain the large distortion and low trapping energies computed for this site and discussed in Section 3.2.2.a.

Analogously, at the other edge site in the fourth layer (**Figure 11d**, Ti$_{5c}^{edge}$) there is an unfavorable starting point for trapping because of a similar compression of the Ti–Ti distance along the left-right direction (3.65 Å) and of two equatorial Ti$_{5c}^{edge}$–O bonds out of three (red).

At the border of the second layer we find O$_{2c}$ atoms instead of the Ti$_{5c}$ (**Figure 11b**). In this case, along the up-down direction, the effect on the "bulk" distances is opposite and the innermost Ti–Ti distance is significantly longer (3.86 Å) than in bulk anatase. This is probably the reason why the hole is better accommodated in O$_{3c}^{core\_ax}$, with higher trapping energy with respect to O$_{3c}^{core\_eq}$ and the bulk O$_{3c}$ site. In the same layer but along the left-right direction, the opposite trend is observed with short Ti–Ti distance (3.73 Å). This is a consequence of the layer termination made up by Ti$_{5c}$ atoms.

In the third layer (**Figure 11c**), a similar but even enhanced behavior than in the second one can be observed, because the opposite facets are closer with shorter Ti-O-Ti chains. In fact, on the "Ti$_{5c}$" (up-down) direction the compression forces the innermost Ti–Ti distance to 3.70 Å, whereas the same distance in the "O$_{2c}$" (left-right) direction is 3.86 Å. In addition, Ti$_{5c}$–Ti$_{5c}$ distances at the border are elongated, because the line terminates with O$_{2c}$ species.

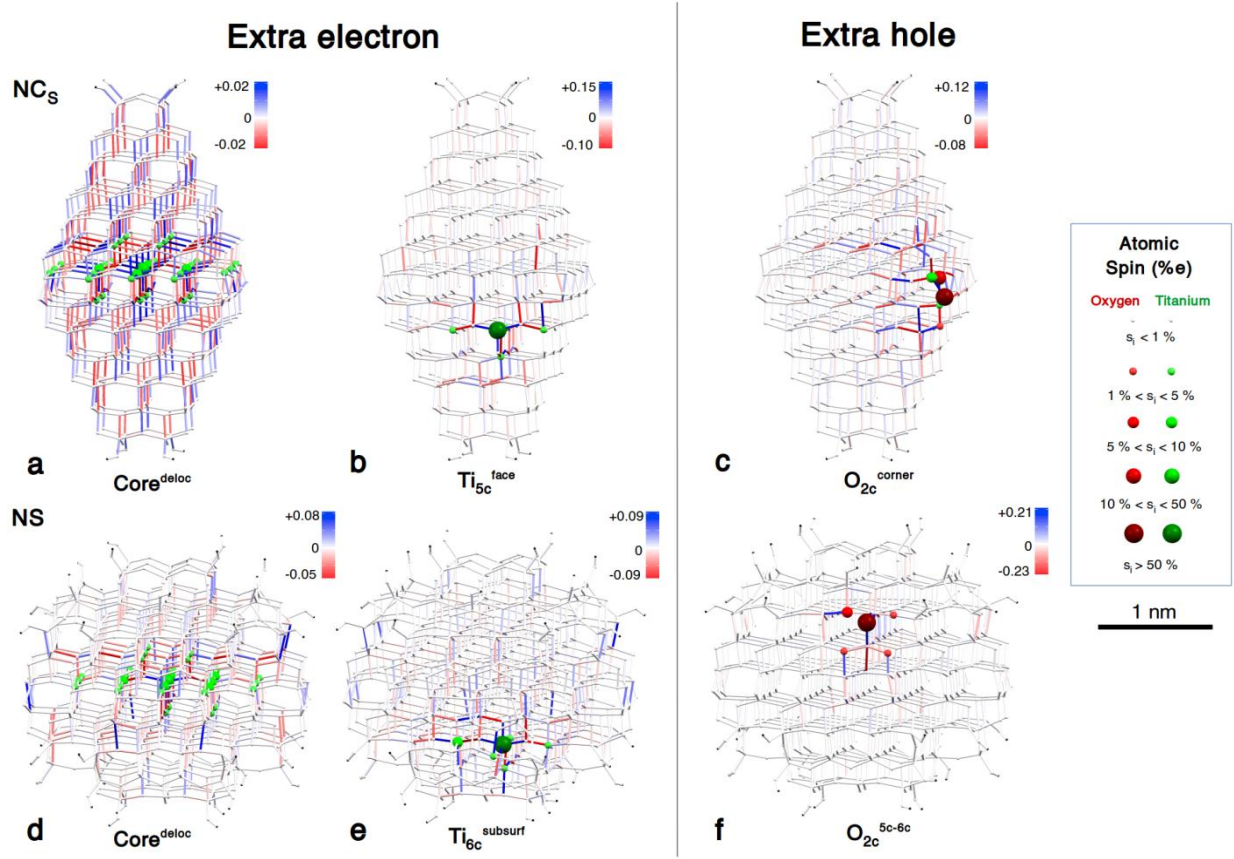

**Fig 12** Visual representation of the elongation/compression of the bonds due to the polaronic distortion associated with the charge trapping in the **NC$_S$** nanocrystal and **NS$_S$** nanosphere. These 3D plots are obtained comparing the neutral B3LYP geometries with the ones with an excess trapped charge. The elongation is given in shade of blue, the compression in shade of red as indicated by the color scale close to each panel. Ti and O atoms are represented as spheres with a radius and color related to the degree of spin localization on each atom, according to the legend (top right). The sites nomenclature is defined graphically in **Figure 1** and **Figure S1** or in the text.

An overview of the polaronic distortions at the electron (**b** and **e**) or hole (**c** and **f**) trapping sites in **NC$_S$** and **NS**, respectively, is graphically represented in **Figure 12**. The delocalized electron solutions are also reported for comparison (**a** and **d**). In these representations, bonds are colored according to the relative elongation (blue)/compression (red) with respect to the relaxed neutral geometry of the nanoparticle, whereas atoms are described by spheres of increasing diameter according to the degree of atomic spin localization.

As far as faceted nanoparticles are concerned, we may observe, as already discussed in Section 3.2.2, that the electron is fully delocalized on the central three x,y layers of the nanoparticle (**a**). However, here we may also notice that the structural distortions involve a large portion of the nanoparticle: in particular all the bonds of the central layers and all the bonds along the vertical or z direction of the nanocrystal, with alternating elongation and compression effects. On the contrary, when the electron is trapped (**b**), the polaronic distortion is much more localized and involves only a portion of the nanocrystal. The radius of this polaron is about 0.5 nm. Analogously, for the trapped hole (**c**), the polaronic distortion involves only a portion of the nanocrystal and the radius of the polaron is below 1 nm in size.

As far as spherical nanoparticles are concerned, the delocalized electron (**d**) is slightly less delocalized than in the faceted nanoparticle: the spin density is not present on the outermost atoms in the three x,y layers of the nanosphere involved in the electron spin hosting. Here the structural distortions are limited to the central three layers and do not involve all the nanoparticle bonds. When the electron is trapped by a Ti center (**e**), only the nearby Ti ions are involved in the trapping process and the polaronic distortion affects the bond lengths within a radius of about 0.5 nm. Finally, in the case of the trapped hole (**f**), the degree of localization is extremely high in terms of spin localization and in terms of structural distortion which involve only the first and second shell arount the trapping O atom, within a radius of less than 0.5 nm.

## 4. Summary and Conclusions

In this work, we have investigated the life path of light-induced energy carriers (excitons) and charge carriers (electrons and holes) in anatase $TiO_2$ nanoparticles of different size and shape by means of a wide set of hybrid density functional theory calculations. It is nowadays well established that a portion of exact exchange in the functional is required in order to account properly for the polaronic distortions[71] accompanying the self-trapping of the carriers at lattice sites in the bulk or at the surface of the nanoparticles. The computational models have been corroborated by an extensive comparison with available experimental data based on photoluminescence measurements, EPR and transient absorption spectroscopies.

As regards excitons in faceted nanoparticles, we have observed that they can become self-trapped, with a highly localized electron on a core $Ti^{3+}$ ion and a highly localized hole on the $O^-$ ion in the axial position ($Ti^{3+}$—$O_{ax}^-$). If the exciton unbinds, various different configurations are possible with the electron and hole in different relative positions, but with similar energy. The most stable configuration is that where both electron and hole have reached the nanoparticles surface. In spherical nanoparticles, the bound exciton trapping energy is smaller (by -0.25 eV), but the degree of localization is still very high. Again, when the carriers separate, they prefer to be at or near the surface, as one would expect. Computed PL emission energies for the self-trapped excitons in **NC** and **NS** are in good agreement with experimental observations.

As regards excess electrons in faceted nanoparticles, rather unexpectedly, we observe that Ti ions at edge and corner sites are not capable of trapping, and that the trapping site $Ti_{5c}$ on (101) facets and $Ti_{6c}$ subsurface have an energy that is comparable with the shallow trap involving several Ti ions in the central layers of the nanocrystal. In spherical nanoparticles, the electron localization is more effective because of a less stable delocalized state on core Ti ions. In excellent agreement with experiments, calculated EPR hyperfine coupling constants are an accurate measure of the charge localization for shallow and deep traps. Computed optical excitations are also in line with TA spectroscopic quantities.

As regards electron holes, we observe that they are preferentially trapped by an $O_{2c}$ ion at a corner site ($Ti_{5c}$-O-$Ti_{4c}$) in faceted nanoparticles and by a surface $O_{2c}$ site ($Ti_{5c}$-O-$Ti_{6c}$) in spherical nanoparticles, where corners do not exist. These two very different species have similar trapping energies (similar IP values), which indicates that spherical nanoparticles are by far more efficient in hole trapping than faceted ones. Isolated OH groups have also been considered as potential hole trapping site but found to be slightly less efficient. In particular, OH groups bound to $Ti_{5c}$ ions are capable of hole trapping with a reduction in trapping energy of 0.1-0.3 eV, depending on the shape of the nanoparticle. In this respect, we are currently investigating the role played by additional water molecules, which would be present in an aqueous medium and would be hydrogen bonded to the pending OH groups covalently bound to the nanoparticles. Their presence could largely affect the hole trapping properties of OH groups.[70] Even more drastic would be the exposure of the photoexcited nanoparticles to molecular oxygen or alcohols, which are excellent electron and hole scavengers, respectively. This, however, will be the subject of a future work.

In the final part of the paper, we present a detailed analysis of how the structural deformations (strain) induced by the nanosize and by the shape affect and control the carriers trapping process. Selective compression or elongation of Ti-O bonds play a key role in determining the most or the least efficient trapping sites.

Before concluding, we wish to comment that the observations reported above can be extended to other nanosystems. However, these must present similar surface-to-bulk ratio and degree of curvature as those of the investigated models.

To conclude, with the present study we have made a step forward in the understanding of energy (excitons) and charge (electrons and holes) carriers in $TiO_2$ nanosized systems with respect to the existing literature, where the attention is still mostly focused on bulk or extended surface $TiO_2$ models. Here, we have shown in detail how $TiO_2$ nanosystems are different when not only size but also shape comes into play. These findings will be very useful for the rationalization of many future experiments based on $TiO_2$ nanostructured materials.

# ASSOCIATED CONTENT
**Supporting Information**

The Supporting Information is available free of charge via the Internet at http://pubs.acs.org.

> Tables of Excitations, PL, Trapping Energies, IP and EA as obtained with HSE06 functional; Histograms of atomic spin distributions for excess electrons and holes; Additional 3D spin density plots of excess electrons and holes in **NC$_S$**, **NC$_L$** and **NS**; Representation of the Ti$_{29}$O$_{58}$ cluster model; Table of hyperfine coupling constants for an excess hole in **NC$_S$** and **NS** as obtained with CRYSTAL14; Projected density of states (PDOS) on Ti t$_{2g}$ and e$_g$ states.

## Acknowledgments

We are grateful to Lorenzo Ferraro and Elisa Albanese for their constant technical help. The project has received funding from the European Research Council (ERC) under the European Union's HORIZON2020 research and innovation programme (ERC Grant Agreement No [647020]).

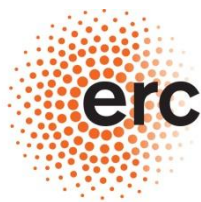

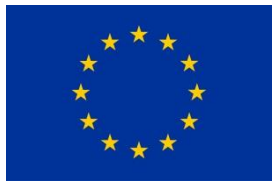

## Biography

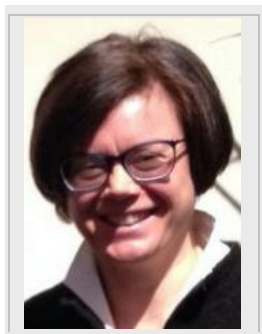

Cristiana Di Valentin graduated in Chemistry (1997) at the University of Pavia where she received her Ph.D. (2000). She was appointed by the University of Milano-Bicocca as Assistant (2002) and Associate Professor (2012). She was visiting scientist at TUM, Universitat Barcelona, Ecole Nationale Superieure Paris and Princeton University. Her research focuses on ab initio study of nanostructured semiconducting oxides and graphene, in (photo)catalysis, fuel cells and nanomedicine. She currently holds an ERC Consolidator Grant

(2016-2021).

Lara Ferrighi graduated in Chemistry at the University of Ferrara (2003). She then moved to the Centre for theoretical and computational Chemistry at Artic University of Norway where she received her PhD. After a postdoc at iNANO-Fysik in Århus (2008-2012), she joined the group of Prof. Di Valentin at the University of Milano-Bicocca where she works as researcher scientist on the FIRB project "Beyond Graphene: tailored C-layers for novel catalytic materials and green chemistry".

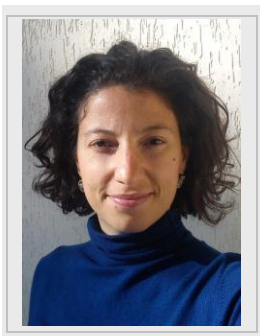

Gianluca Fazio graduated in Chemistry in 2014 at the University of Milano-Bicocca, with a thesis on doped graphene in fuel cells. He is now pursing his Ph.D. degree under the supervision of Prof. Di Valentin. He has already received several prizes: 'Fondazione Grazioli' prize, Master's Degree prize by Electrochemical division of Italian Chemical Society and he was a Young Scientist at the 65th Lindau Nobel Meeting. His current research focuses on modelling semiconductive oxide nanoparticles.

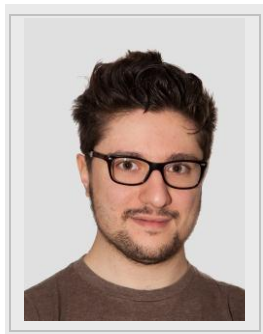

## References

[1] Y. Bai, I. Mora-Seró, F. De Angelis, J. Bisquert, P. Wang, Chem. Rev. 114 (2014) 10095–10130.

[2] Y. Ma, X. Wang, Y. Jia, X. Chen, H. Han, C. Li, Chem. Rev. 114 (2014) 9987–10043.

[3] J. Schneider, M. Matsuoka, M. Takeuchi, J. Zhang, Y. Horiuchi, M. Anpo, D.W. Bahnemann, Chem. Rev. 114 (2014) 9919–9986.

[4] M. Kapilashrami, Y. Zhang, Y.-S. Liu, A. Hagfeldt, J. Guo, Chem. Rev. 114 (2014) 9662–9707.

[5] P. Szymanski, M. A. El-Sayed, Theor. Chem. Acc. 131 (2012) 1202 – 1225.

[6] L. Sang, Y. Zhao, C. Burda, Chem. Rev. 114 (2014) 9283 – 9318.

[7] M. A. El-Sayed, Acc. Chem. Res. 37 (2004) 326 – 333.

[8] S. W. Koch, M. Kira, G. Khitrova, H. M. Gibbs, Nat. Mater. 5 (2006) 523 – 531.

[9] H. Tang, H. Berger, P.E. Schmid, F.Lévy, Solid State Commun. 87 (1993) 847 – 850.

[10] G. Wannier, Physical Review 52 (1937) 191.

[11] N. F. Mott, A. M. Stoneham, J. Phys. C: Solid State Phys. 10 (1977) 3391.

[12] C. Richter, C. A. Schmuttenmaer, Nature Nanotech. 5, 2010, 769 – 772.

[13] Y.-F. Li, Z.-P. Liu, J. Am. Chem. Soc. 133 (2011) 15743 – 15752.

[14] V. Luca, J. Phys. Chem. C 113 (2009) 6367 – 6380.

[15] C. Di Valentin, G. Pacchioni, A. Selloni, Phys. Rev. Lett. 97 (2006) 166803.

[16] C. Di Valentin, G. Pacchioni, A. Selloni, J. Phys. Chem. C 113 (2009) 20543 – 20552.

[17] N. A. Deskin, R. Rousseau, M. Dupuis, J. Phys. Chem. C 113 (2009) 14583 – 14586.

[18] S. Kerisit, N. A. Deskins, K. M. Rosso, M. A. Dupuis, J. Phys. Chem. C 112 (2008) 7678 – 7688.

[19] C. Burda, X. Chen, R. Narayanan, M. A. El-Sayed, Chem. Rev. 105 (2005) 1025 – 1102.

[20] R. Penn, J. F. Banfield, Geochim. Cosmochim. Acta 63 (1999) 1549 – 1557.

[21] M. Lazzeri, A. Vittadini, A. Selloni, Phys. Rev. B 63 (2001) 155409.

[22] T. Rajh, N. M. Dimitrijevic, M. Bissonnette, T. Koritarov, V. Konda, Chem. Rev. 114 (2014) 10177 – 10216.

[23] R. Dovesi, V. R. Saunders, C. Roetti, R. Orlando, C. M. Zicovich-Wilson, F. Pascale, B. Civalleri, K. Doll, N. M. Harrison, I. J. Bush, P. D'Arco, M. Llunell, M. Causà and Y. Noël CRYSTAL14, (2014) CRYSTAL14 User's Manual. University of Torino, Torino.

[24] P. J. Stephens, F. J. Devlin, C. F. Chabalowski, M. J. Frisch, J. Phys. Chem. 98 (1994) 11623-11627.

[25] A. D. Becke J. Chem. Phys. 98 (1993) 5648-5652.

[26] A.V. Krukau, O.A. Vydrov, A. F. Izmaylov, G. E. Scuseria, J. Chem. Phys. 125 (2006) 224106.

[27] J. K. Burdett, T. Hughbanks, G.J. Miller, J. W. Richardson, J. V. Smith, J. Am. Chem. Soc. 109 (1987) 3639-3646.

[28] G. Fazio, L. Ferrighi, C. Di Valentin, J. Phys. Chem. C 119 (2015) 20735 – 20746.

[29] A. S. Barnard, P. Zapol, Phys. Rev. B: Condens. Matter Mater. Phys 70 (2004) 235403.

[30] B. Civalleri, Ph. D'Arco, R. Orlando, V.R. Saunders, R. Dovesi, Chem. Phys. Lett. 348 (2001) 131.

[31] M. J. Frisch, G. W. Trucks, H. B. Schlegel, G. E. Scuseria, M. A. Robb, J. R. Cheeseman, G. Scalmani, V. Barone, B. Mennucci, G. A. Petersson et al. Gaussian, Inc. (2009) Wallingford CT.

[32] F. Gallino, G. Pacchioni, C. Di Valentin, J. Chem. Phys. 133 (2010) 144512.

[33] S. Lany, A. Zunger, Phys. Rev. B 78 (2008) 235104.

[34] M. Watanabe, T. Hayashi, J. Lumin. 112 (2005) 88 – 91.

[35] H. Tang, K. Prasad, R. Sanjinès, P. E. Schmid, F. Lévy, J. Appl. Phys. 75 (1994) 2042.

[36] H. Tang, H. Berger, P. E. Schmid, F. Levy, Solid State Commun. 92 (1994) 267 – 271.

[37] H. Najafov, S. Tokita, S. Ohshio, A. Kato, H.Saitoh, Jpn. J. Appl. Phys. 44 (2005) 245 – 253.

[38] T. Sekiya, S. Kamei, S. Kurita, J. Lumin. 87-89 (2000) 1140 – 1142.

[39] H. Bieber, P. Gilliot, M. Gallart, N. Keller, V. Keller, S. Begin-Colin, C. Pighini, N. Millot, Catal. Today 122 (2007) 101 – 108.

[40] W. F. Zhang, M.S. Zhang, Z. Yin, Q. Chen, Appl. Phys B 70 (2000) 261 – 265.

[41] L. Cavigli, F. Bogani, A. Vinattieri, V. Faso, G. Baldi, J. Appl. Phys. 106 (2009) 053516.

[42] L. Cavigli, F. Bogani, A. Vinattieri, L. Cortese, M. Colucci, V. Faso, G. Baldi, Solid State Sci. 12 (2010) 1877.

[43] Y. Liu, R. O. Claus, J. Am. Chem. Soc. 119 (1997) 5273 – 5274.

[44] D. Pan, N. Zhao, Q. Wang, S. Jiang, X. Ji, L. An, Adv. Mater. 17 (2005) 1991 – 1995.

[45] A. M. Peiro, J. Peral, C. Domingo, X. Domeneck, J. A. Ayllon, Chem. Mater. 13 (2001) 2567 – 2573.

[46] K. Ding, Z. Miao, Z. Liu, Z. Zhang, B. Han, G. An, S. Miao, Y. Xie, J. Am. Chem. Soc. 129 (2007) 6362 – 6363.

[47] C. Di Valentin, A. Selloni, J. Phys. Chem. Lett. 2 (2011) 2223 – 2228.

[48] J. Muscat, J. Swamy, N. M. Harrison, Phys. Rev. B 65 (2002) 224112.

[49] L. Chiodo. J. M. García-Lastra, A. Iacomino, S. Ossicini, J. Zhao, H. Petek, A. Rubio, Phys. Rev. B 82 (2010) 045207.

[50] D. A. Panayotov, J. T. Yates Jr, Chem. Phys. Lett. 436 (2007) 204.

[51] A.Yamakata, T. Ishibashi, H. Onishi, Chem. Phys. Lett. 333 (2001) 271 – 277.

[52] J. R. Durrant, J. Photochem. Photobiol. A 148 (2002) 5 – 10.

[53] S. T. Martin, H. Hermann, M. R. Hoffmann, J. Chem. Soc., Faraday Trans. 90 (1994) 3323 – 3330.

[54] N. Beermann, G. Boschloo, A. Hagfeldt, J. Photochem. Photobiol. A 152 (2002) 213.

[55] G. Boschloo, D. Fitzmaurice, J. Phys. Chem. B 103 (1999) 2228 – 2231.

[56] S. H. Szezepankiewicz, J. A. Moss, M. R. Hoffmann, J. Phys. Chem. B 106 (2002) 2922 – 2927.

[57] T. Berger, M. Sterrer, O. Diwald, E. Knözinger, ChemPhysChem 6 (2005) 2104 – 2112.

[58] T. Berger, M. Sterrer, S. Stankic, J. Bernardi, O. Diwald, E. Knözinger, Mater. Sci. Eng. C 25 (2005) 664 – 668.

[59] M. Chiesa, M. C. Paganini, S. Livraghi, E. Giamello, Phys. Chem. Chem. Phys. 15 (2013) 9435 – 9447.

[60] O. I. Micic, Y. Zhang, K. R. Cromack, A. D. Trifunac, M. C. Thurnauer, J. Phys. Chem. 97 (1997) 7227 – 7283.

[61] V. Brezová, Z. Barbieriková, M. Zukalová, D. Dvoranová, L. Kavan, Catal. Today 230 (2014) 112 – 118.

[62] L. J. Antila, F. G. Santomauro, L. Hammarström, D. L. A. Fernandes, J. Sá, Chem. Commun. 51 (2015) 10914 – 10916.

[63] S. Moser, L. Moreschini, J. Jacimovic, O. S. Barisic, H. Berger, A. Magrez, Y. J. Chang, K. S. Kim, A. Bostwick, E. Rotenberg, L. Forro, M. Grioni, Phys. Rev. Lett. 110 (2013) 196403.

[64] T. Yoshihara, R. Katoh, A. Furube, Y. Tamaki, M. Murai, K. Hara, S. Murata, H. Arakawa, M. Tachiya, J. Phys. Chem. B 108 (2004) 3817–3823.

[65] I. A. Shkrob, M. C. Sauer Jr., J. Phys. Chem. B 108 (2004) 12497 – 12511.

[66] M. H. Rittmann-Frank, C. J. Milne, J. Rittmann, M. Reinhard, T. J. Penfold, M. Chergui, Angew. Chem. Int. Ed. 53 (2014) 5858 – 5862.

[67] Z. Zhang, T. F. Hughes, M. Steigerwald, L. Brus, R. A. Freisner, J. Am. Chem. Soc. 134 (2012) 12028 – 12042.

[68] S. Maurelli, S. Livraghi, M. Chiesa, E. Giamello, S. Van Doorslaer, C. Di Valentin, G. Pacchioni, Inorg. Chem. 50 (2011) 2385 – 2394.

[69] N. A. Deskins, M. Dupuis, J. Phys. Chem. C 113 (2009) 346 – 358.

[70] K. Shirai, T. Sugimoto, K. Watanabe, M. Haruta, H. Kurata, Y. Matsumoto Nano Lett. 16 (2016) 1323 – 1327.

[71] F. De Angelis, C. Di Valentin, S. Fantacci, A. Vittadini, A. Selloni, Chem. Rev. 114 (2014) 9708 – 9753.

**Highlights**

- Energy (excitons) and charge ($e^-/h^+$) carriers in ~3 nm anatase $TiO_2$ NPs are modeled;
- Accurate first-principles calculations can describe polarons and polaron pairs in NPs;
- Strain induced by nanosized dimensions and curvature controls carriers self-trapping;
- Spherical and faceted nanoparticles behave differently;
- Computed quantities agree with experimental PL, EPR and TA spectroscopic data.

Graphical Abstract

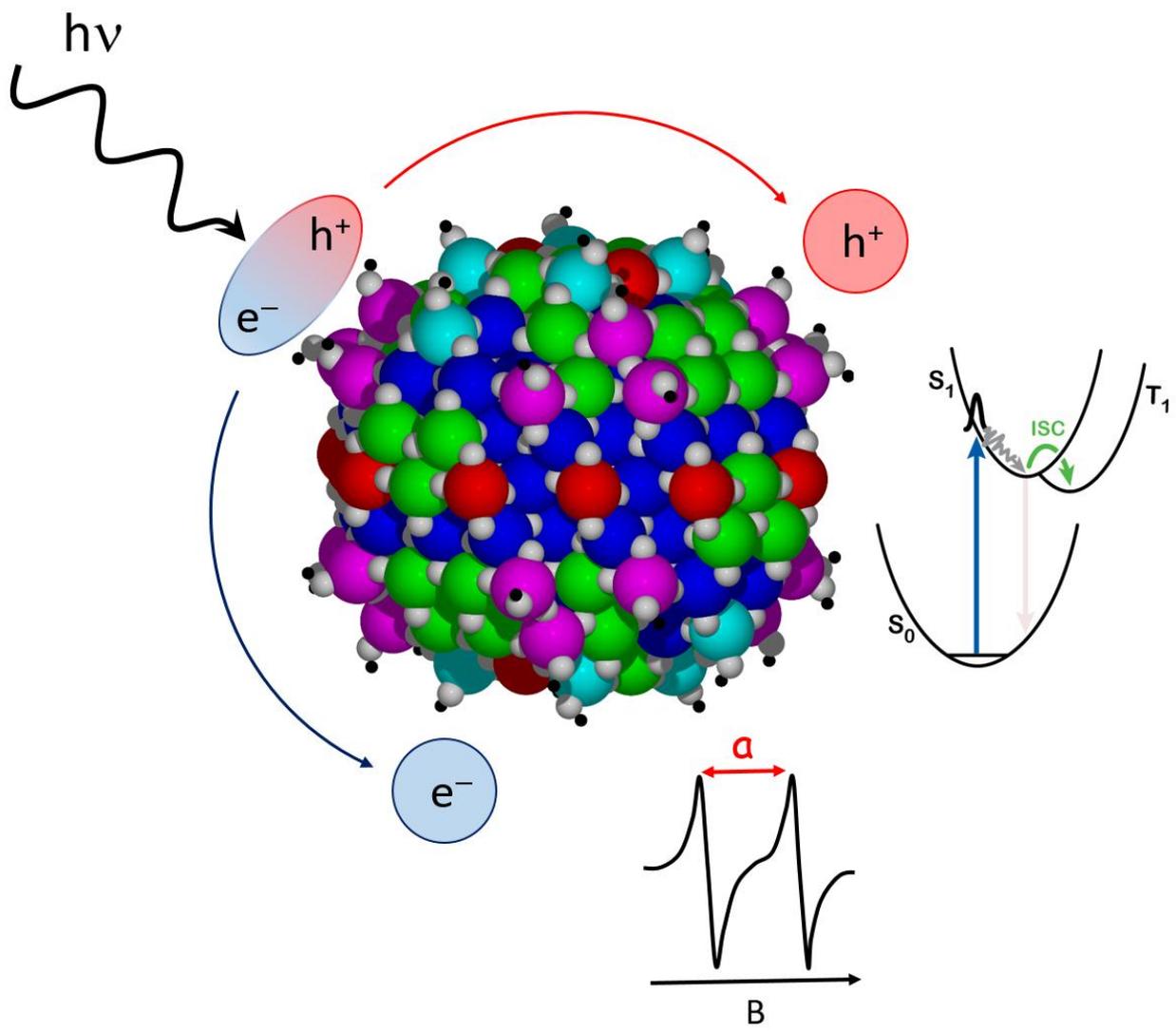

**Supporting Information for**

**PHOTOEXCITED CARRIERS RECOMBINATION AND TRAPPING**

**IN SPHERICAL VS FACETED TiO$_2$ NANOPARTICLES**

Gianluca Fazio, Lara Ferrighi, Cristiana Di Valentin[*]

Dipartimento di Scienza dei Materiali, Università di Milano-Bicocca,

via R. Cozzi 55, 20125 Milano Italy

---

[*] Corresponding author: cristiana.divalentin@mater.unimib.it

**Table S1** Singlet-Triplet Vertical Excitation Energy, Trapping Energy of the Triplet Exciton and its calculated photoluminescence (PL) in the Axial configuration for **NC$_S$** nanocrystal as obtained with HSE06 functional. All energies are in eV.

|  |  | NC$_S$ |
|---|---|---|
| ($S_0 \rightarrow T_1$)$_{vert}$ | | 3.87 |
| $\Delta E_{trap}$ | Ti$^{3+}$ – O$_{ax}^{-}$ | -0.82 |
| PL | | 1.68 |

**Table S2** Trapping Energy (ΔE) of the triplet exciton at different sites with the charge localization (%electron or %hole) in **NC$_S$** and **NS$_S$** anatase nanoparticles with the HSE06 functionals. No symmetry constrains are imposed to all the calculations. Energies are in eV. The sites nomenclature is defined graphically in **Figure 1**.

| Model | Position | ΔE | %electron | %hole |
|---|---|---|---|---|
| \multicolumn{5}{c}{Electron/Hole pairs in **NC$_S$**} | | | | |
| b | Ti$_{6c}^{core}$ – O$_{3c}^{core}$ (Ti$^{3+}$ – O$_{ax}^{-}$) | -0.82 | 6% | 93% |
| d | Ti$_{6c}^{core}$ – O$_{2c}^{corner}$ | -0.82 | 6% | 73% |
| e | Ti$_{5c}^{face}$ – O$_{2c}^{corner}$ | -0.82 | 92% | 73% |
| \multicolumn{5}{c}{Electron/Hole pairs in **NC$_L$**} | | | | |
| | Ti$_{6c}^{core}$ – O$_{3c}^{core}$ (Ti$^{3+}$ – O$_{ax}^{-}$) | -0.82 | 6% | 93% |

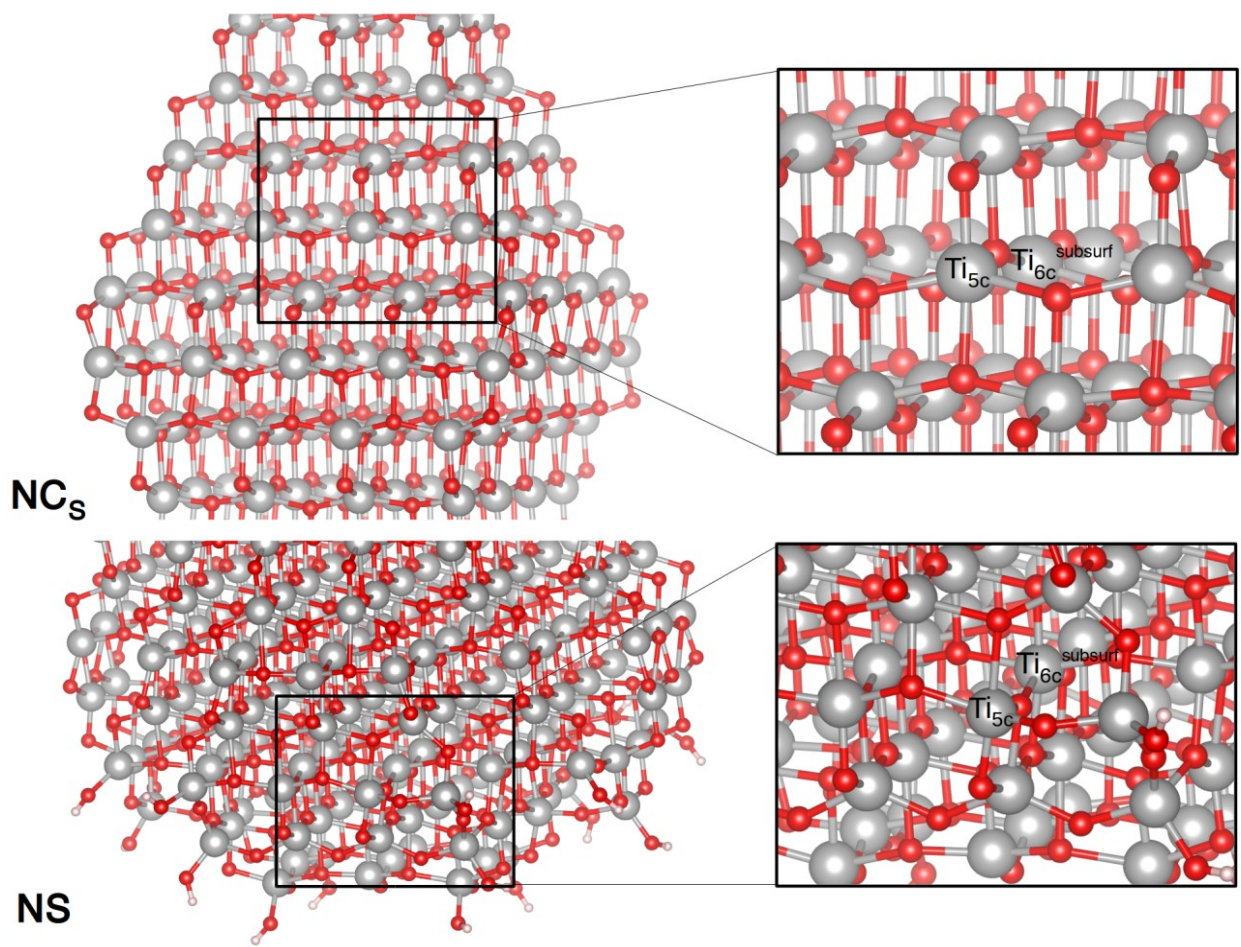

**Figure S1:** Position of the Ti atoms defined as subsurface $Ti_{6c}$ ($Ti_{6c}^{subsurf}$) in the **NC_S** nanocrystal (top panel) and **NS** nanosphere (bottom panel).

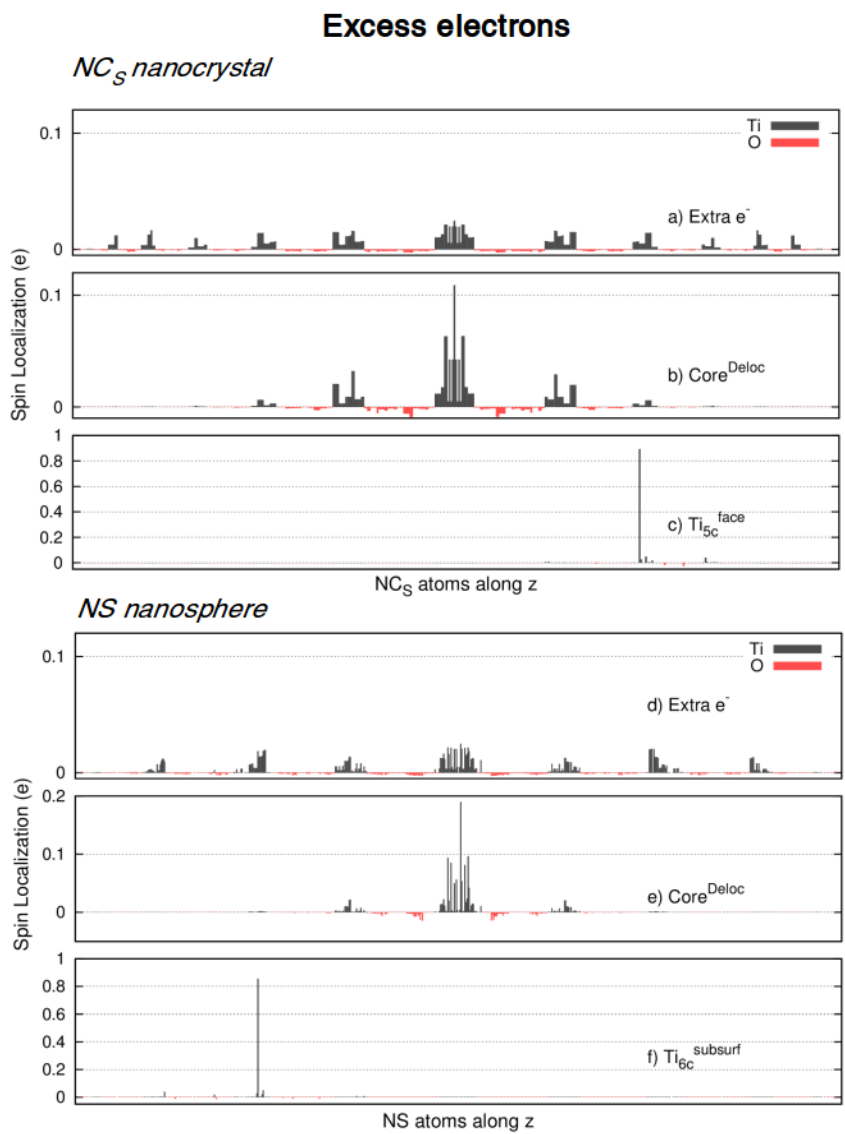

**Figure S2:** Histograms of the atomic spin on each oxygen or titanium atoms in the **NC$_S$** (top) and **NS** (bottom) nanoparticles an excess electron. The distribution is given for three cases: a) when the charge is added vertically to the ground state geometry, b) when the geometry is allowed to relax with an excess charge, c) in the best trapping site for the charge. Atoms are ordered from bottom to top by their z coordinate. The sites nomenclature is defined graphically in **Figure 1** and **Figure S1** or in the text.

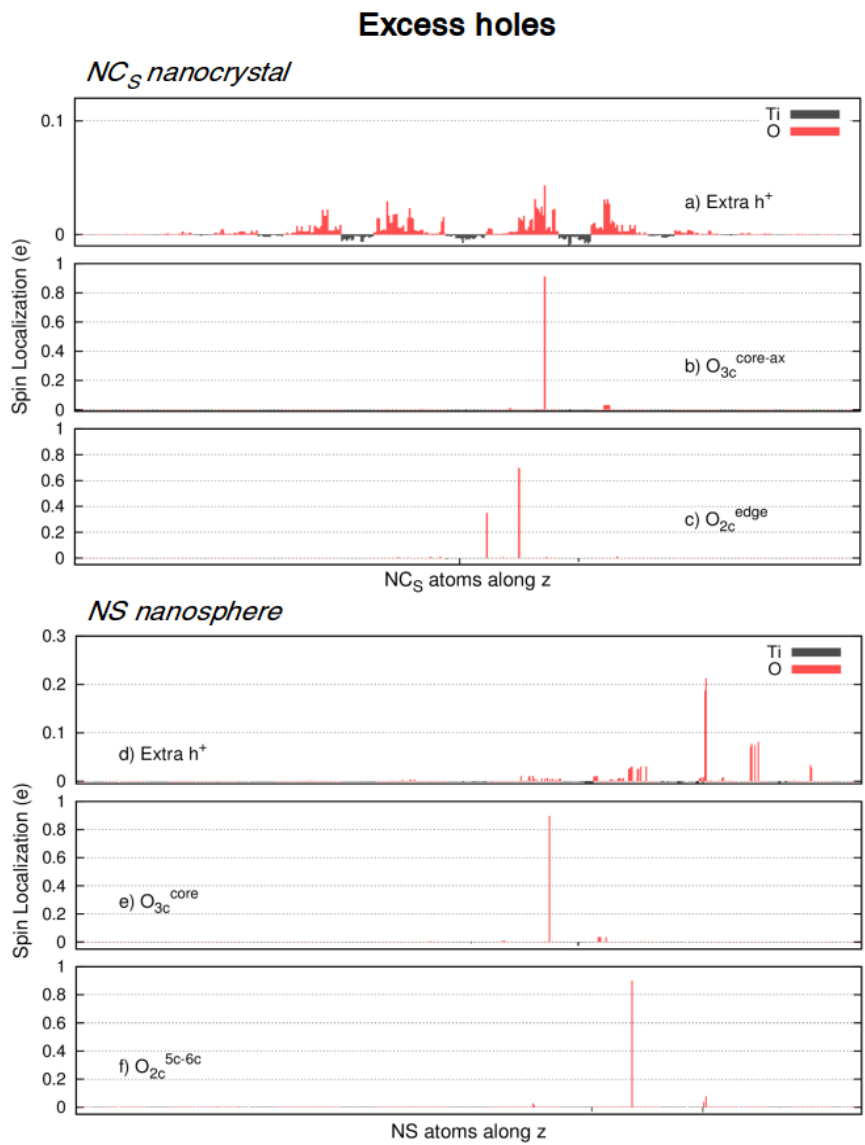

**Figure S3:** Histograms of the atomic spin on each oxygen or titanium atoms in the **NC_S** (top) and **NS** (bottom) nanoparticles an excess hole. The distribution is given for three cases: a) when the charge is added vertically to the ground state geometry, b) when the geometry is allowed to relax with an excess charge, c) in the best trapping site for the charge. Atoms are ordered from bottom to top by their z coordinate. The sites nomenclature is defined graphically in **Figure 1** or in the text.

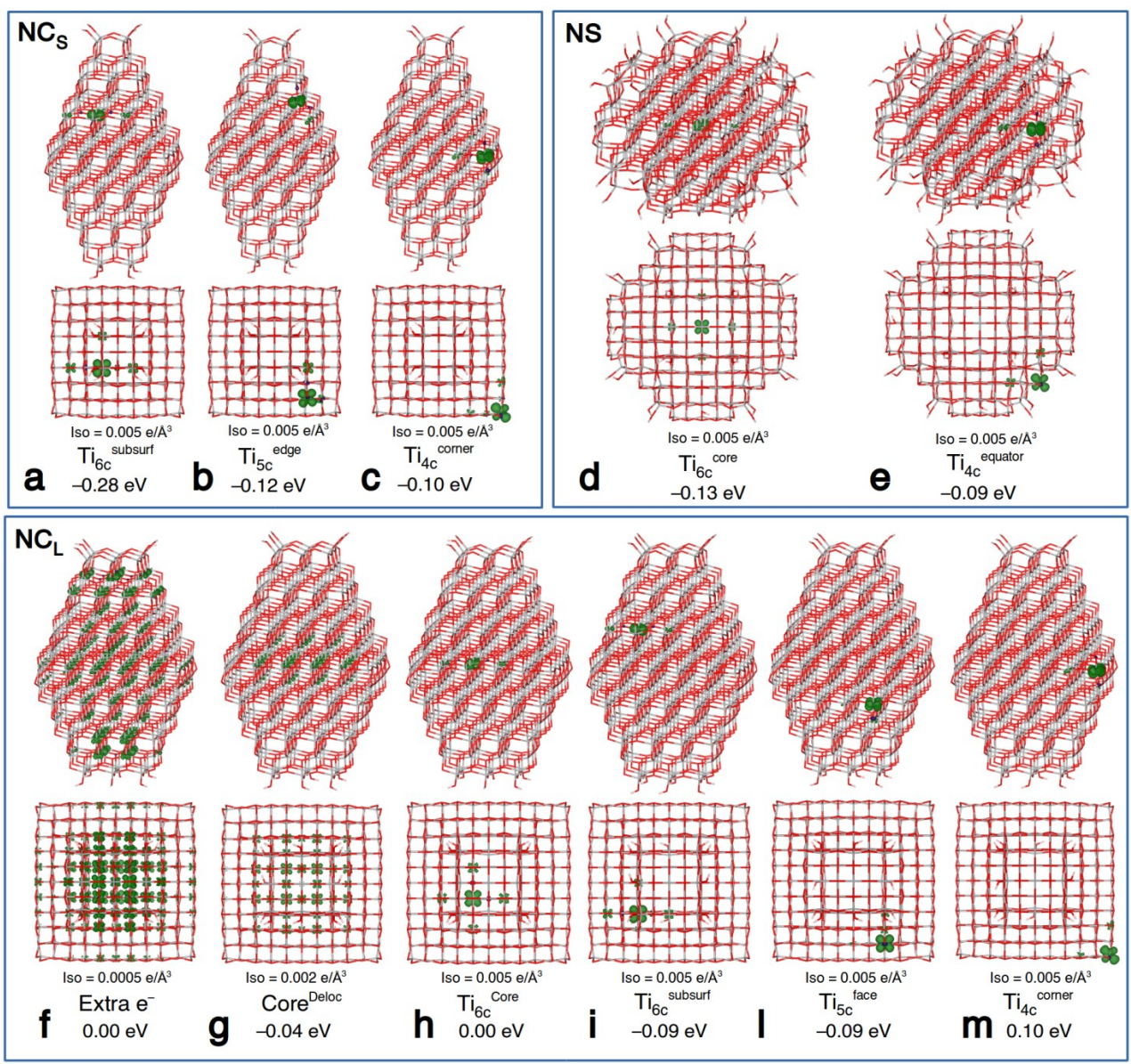

**Figure S4** Additional front and top views of the 3D plots of spin density of trapped electrons in **NC$_S$** and **NC$_L$** nanocrystals (top panel left and bottom panel, respectively) and in **NS** nanosphere (top panel right). Below each structure the isovalue of each 3D plot and the energy gain ($\Delta E_{trap}$) relative to the vertical addition of an excess charge are given. The sites nomenclature is defined graphically in **Figure 1** and **Figure S1** or in the text.

**Table S3** Electron Affinity (EA) and Trapping Energy ($\Delta E_{trap}$) for electrons at different sites for the three anatase nanoparticles with HSE06 functionals. The reference zero for $\Delta E_{trap}$ is obtained by removing one electron with no atomic relaxation. The charge localization in % of electron is also given. No symmetry constrains are imposed in the calculations. Energies are in eV. The sites nomenclature is defined graphically in **Figure 1** and **Figure S1** or in the text.

| Position | $\Delta E_{trap}$ | Localization | EA |
|---|---|---|---|
| Excess electron in **NC$_S$** | | | |
| Vertical | | | -3.41 |
| $Ti_{4c}^{corner}$ | 0.00 | 92% | -3.41 |
| $Ti_{5c}^{face}$ | -0.14 | 92% | -3.55 |
| $Ti_{5c}^{edge}$ | 0.02 | 91% | -3.39 |
| $Ti_{6c}^{subsurf}$ | -0.11 | 77% | -3.52 |
| $Core^{deloc}$ | -0.10 | 5% | -3.51 |

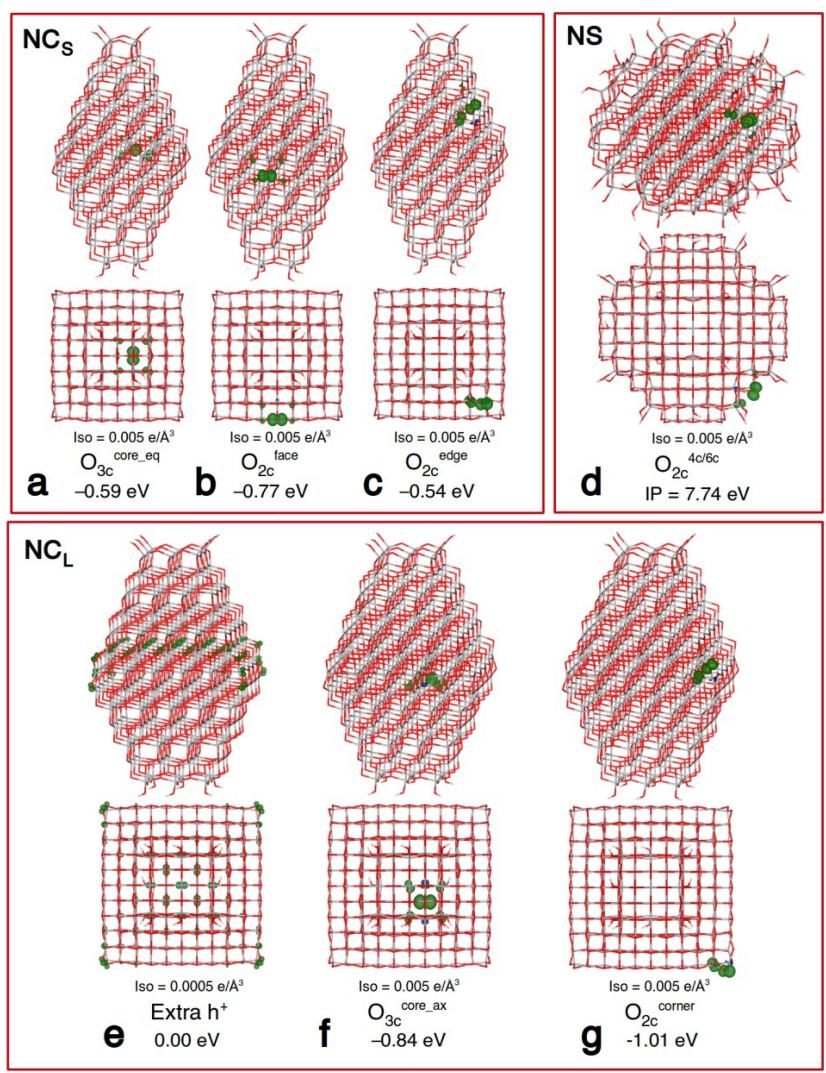

**Figure S5** Additional front and top views of the 3D plots of spin density of trapped holes in **NC$_S$** and **NC$_L$** nanocrystals (top panel left and bottom panel, respectively) and in **NS** nanosphere (top panel right). Below each structure the isovalue of each 3D plot and the energy gain ($\Delta E_{trap}$) relative to the vertical addition of an excess charge are given. The sites nomenclature is defined graphically in **Figure 1** or in the text.

**Table S4** Ionization Potential (IP), Trapping Energy ($\Delta E_{trap}$) for an excess hole at different sites for the three anatase nanoparticles with the HSE06 functional. The reference zero for $\Delta E_{trap}$ is obtained by removing one electron with no atomic relaxation. Their charge localization in % of electron is also given. No symmetry constrains are imposed in all the calculations. Energies are in eV. The sites nomenclature is defined graphically in **Figure 1** or in the text.

| Position | $\Delta E_{trap}$ | Localization | IP |
|---|---|---|---|
| Hole addition in **NC$_S$** | | | |
| Vertical | | | 8.93 |
| $O_{2c}^{corner}$ | -0.96 | 69%/31% | 7.97 |
| $O_{3c}^{core\_ax}$ | -1.06 | 93% | 7.87 |

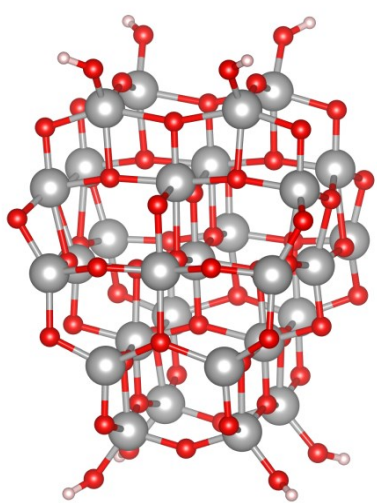

**Figure S6.** Ball-and-Stick representation of the $Ti_{29}O_{58} \cdot 4\ H_2O$ cluster used for g-tensor calculations.

**Table S5** Hyperfine coupling constants (hpcc, in G) tensor with $^{17}O$, decomposed as the Fermi contact term ($a_{iso}$) and the dipolar tensor (B) for an excess hole at the for various hole trapping sites in **NC$_S$** and **NS**, as calculated with the CRYSTAL14 code. The sites nomenclature is defined graphically in Errore. L'origine riferimento non è stata trovata. or in the text.

| Position | $a_{iso}$ | $B_{aa}$ | $B_{bb}$ | $B_{cc}$ |
|---|---|---|---|---|
| A-tensor in **NC$_S$** | | | | |
| **$O_{2c}^{face}$** | **-36** | **-91** | **45** | **45** |
| $O_{2c}^{corner}$ | -30 | -36 | 72 | -36 |
| $O_{2c}^{edge}$ | -30 | -38 | 75 | -38 |
| $O_{3c}^{core\_ax}$ | -38 | -90 | 45 | 45 |
| $O_{3c}^{core\_eq}$ | -36 | 42 | -84 | 42 |
| A-tensor in **NS** | | | | |
| $O_{2c}^{4c-6c}$ | -35 | -87 | 43 | 44 |
| **$O_{2c}^{5c-6c}$** | **-35** | **44** | **-88** | **44** |
| $O_{3c}^{core\_ax}$ | -38 | -89 | 44 | 44 |

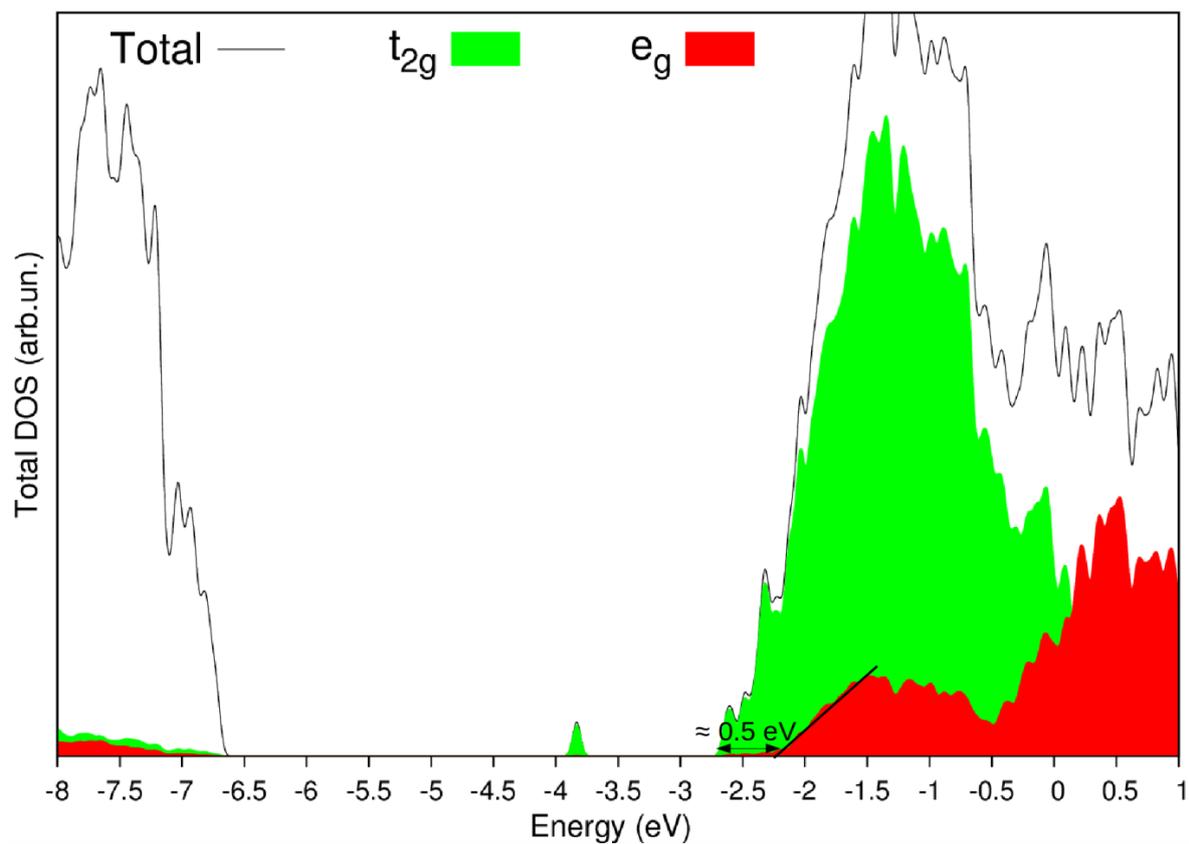

**Figure S7.** Total (DOS) and projected (PDOS) density of states on Ti $t_{2g}$ and $e_g$ states of the **NC$_S$** nanocrystal with an excess electron in Ti$_{5c}^{face}$ site. The black line indicates the inset of the Ti $e_g$ band, which starts around 0.5 eV above the bottom of the conduction band.